\documentclass[%
reprint,
superscriptaddress,
 amsmath,amssymb,
 aps,
]{revtex4-2}
\usepackage[colorlinks=true, allcolors=blue]{hyperref}
\usepackage{graphicx}
\usepackage{bm}
\usepackage[mathlines]{lineno}
\usepackage{multirow}

\usepackage{xr}
\usepackage{orcidlink}

\newcommand{\SPINTEC}{Universit\'e Grenoble Alpes, CEA, CNRS, Grenoble INP, SPINTEC, 38000 Grenoble, France} 
\newcommand{\ANKARA}{Department of Physics, Faculty of Science, Ankara University, 06100, Ankara, Turkey} 
\newcommand{\Handai}{The University of Osaka, Toyonaka, Osaka 560-8531, Japan}
\newcommand{\iuf}{Institut universitaire de France, 75321 Paris, France}

\begin{document}

\preprint{APS/123-QED}

\title{Frustrated out-of-plane Dzyaloshinskii-Moriya interaction \\ and  the onset of atomic-scale 3$q$ magnetic textures \\ in 2D Fe$_{3}$GeXTe (X = Te, Se, S) monolayers}

\author{Rabia Caglayan\orcidlink{0000-0002-4373-064X}}
\thanks{R. C. and L. D. contributed equally.} 
\email{rcaglayan@ankara.edu.tr}
\affiliation{Department of Physics, Graduate School of Natural and Applied Sciences, Ankara University, 06110, Ankara, Turkey}

\author{Louise Desplat\orcidlink{0000-0002-8968-1046}} 
\thanks{R. C. and L. D. contributed equally.} 
\email{louise.desplat@cea.fr}
\affiliation{\SPINTEC}

\author{Sergey Nikolaev\orcidlink{}}
\affiliation{\Handai}

\author{Fatima Ibrahim\orcidlink{}}
\affiliation{\SPINTEC}

\author{Jing Li\orcidlink{0000-0002-9283-3821}}
\affiliation{Universit\'e Grenoble Alpes, CEA, Leti, F-38000 Grenoble, France}

\author{Yesim Mogulkoc\orcidlink{0000-0002-7502-1522}}
\affiliation{Department of Physics Engineering, Faculty of Engineering, Ankara University, 06100, Ankara, Turkey}

\author{Aybey Mogulkoc\orcidlink{0000-0002-0119-4156}}
\affiliation{\ANKARA}

\author{Mairbek Chshiev\orcidlink{0000-0001-9232-7622}}
\affiliation{\SPINTEC}
\affiliation{\iuf}

\date{\today}

\begin{abstract}
We theoretically study the effect of in- and out-of-plane Dzyaloshinskii-Moriya interaction (DMI) on the magnetic ground states of two-dimensional (2D) Fe$_3$GeXTe (X=Te, Se, S) monolayers, where X=Se, S correspond to antisymmetric Janus structures with nonvanishing in-plane DMI. We perform atomistic spin simulations with the extended Heisenberg Hamiltonian parametrized by first principles calculations. While we find that the base DMI in all systems is too weak to stabilize noncollinear states, we show how the frustrated out-of-plane DMI tends to favor atomic-scale $3q$ magnetic textures at the edge of the Brillouin zone. Owing to the ability to tune the DMI in 2D magnets via applied strain or electric field, we study the evolution of the systems' ground state with increasing DMI amplitude. We find that nonplanar $3q$ states are favored under scaling factors as low as 3, while larger DMI tends to stabilize states reminiscent of nanoskyrmion lattices at the atomic-scale.
\end{abstract}

\maketitle

\section{Introduction}

The recent discovery of long-range ferromagnetic order in 2D materials~\cite{gong2017discovery,huang2017layer,deng2018gate} has opened a new playground for 2D spintronics, offering potential for ultra-compact devices with unprecedented material engineering capabilities, and providing a fertile ground for studying exotic spin phenomena~\cite{Wang_acsnano.1c09150,yang_two-dimensional_2022}.

A currently promising entity in the field of spintronics is the magnetic skyrmion~\cite{bogdanov1989thermodynamically,bogdanov1994thermodynamically}, a particle-like, topologically nontrivial spin texture at the nanoscale. Skyrmions are typically stabilized by the interplay of magnetic interactions in the presence of the chiral Dzyaloshinskii-Moriya interaction~\cite{dzyaloshinskii,moriya}, arising under a combination of broken inversion symmetry and strong spin-orbit coupling~\cite{Fert_et_al_DMI_Review_JPSJ.92.081001}. Envisioned as novel information carriers~\cite{fert2013skyrmions},  skyrmions have been extensively studied in bulk magnets~\cite{muhlbauer2009skyrmion,yu2010real}, and transition metal thin films and multilayers~\cite{romming2013writing,fert2013skyrmions,moreau2016additive,boulle_room-temperature_2016}. More recently, skyrmions and other topological textures were also observed in 2D materials ~\cite{ding2019observation,wu2020neel,park2021neel,han2019,khela2023,10.1038/s41467-024-47579-9,doi:10.1021/acsnano.4c00853,PhysRevB.108.214417}.

Amongst the rising number of reported 2D magnets, Fe$_3$GeTe$_2$ (FGT$_{2}$) stands out as a strong candidate for 2D spintronics, thanks to its perpendicular magnetic anisotropy, a high Curie temperature--demonstrated to be gate-tunable up to near-room temperature~\cite{deng2018gate}--and its intriguing charge transport properties~\cite{xu2019large}.
Experimental evidence of magnetic skyrmions was reported in
 FGT$_{2}$ flakes~\cite{ding2019observation,wu2020neel,park2021neel,chakraborty2022magnetic,wu2022a,birch2022history}, 
where their stabilization was attributed either to dipolar interactions~\cite{ding2019observation,birch2022history}, or to DMI induced by the oxidized interface~\cite{park2021neel} or defects~\cite{chakraborty2022magnetic}. It was also proposed that fourth-order interactions, rather than the DMI, may be responsible for stabilizing topological spin textures in FGT$_{2}$~\cite{xu2022assembling}.
Indeed, as in many 2D magnets, uncompensated in-plane DMI is forbidden in pristine FGT$_{2}$, due to the out-of-plane mirror symmetry~\cite{moriya}. Instead, DMI in FGT$_{2}$ was predicted to be predominantly out-of-plane~\cite{laref2020elusive}.  Such a term favors in-plane rotating N\'eel-type spin spirals, meaning that it cannot stabilize traditional skyrmions; it was also shown to have a negligible influence on the formation of skyrmions stabilized via in-plane DMI~\cite{liang2020very,du2022spontaneous}. It is therefore typically neglected in theoretical studies~\cite{shen2022strain,du2022spontaneous,li2023tuning,li2024stability}.
Additional quenching of the out-of-plane DMI close to the center of the first Brillouin zone was also predicted~\cite{laref2020elusive}, implying that it is unlikely to stabilize chiral textures in the long wavelength limit. Instead, it was suggested to induce canted magnetic textures at the boundary of nanoflakes or nanoribbons, and potentially stabilize homochiral planar structures at short wavelengths~\cite{laref2020elusive}. 

Nevertheless, in-plane DMI can be induced in van der Waals heterostructures by proximity effects~\cite{sun2020controlling,sun2021manipulation}. This includes FGT$_{2}$-based systems~\cite{li2023tuning,li2024stability}, where skyrmions and other chiral textures--such as bimerons--have been experimentally observed~\cite{yang2020creation,wu2022a} or theoretically predicted~\cite{li2024stability,goerzen2024emergence}.
Another promising route is Janus structures~\cite{liang2020very,yuan2020intrinsic,xu2020topological, jiang2021topological,cui2020strain,zhang2020emergence,du2022spontaneous,PhysRevB.110.094440,hnw5-nkzh}, where the out-of-plane symmetry is broken by using different elements in the top and bottom layers, with the possibility to tune the DMI amplitude via mechanical strain~\cite{cui2020strain,shen2022strain} or out-of-plane electric fields~\cite{liu2018analysis,liu2018electrical,behera2019magnetic}.
Alternatively, interfacing 2D magnets with a ferroelectric material in multiferroic heterostructures can enable the control of DMI through ferroelectric polarization~\cite{li2021writing,huang2022ferroelectric}.

In this work, we theoretically study the effect of in-plane and out-of-plane DMI in 2D Fe$_{3}$GeXTe  (FGTX, X=Te, Se, S) monolayers, where X = S, Se correspond to Janus structures. We use density functional theory (DFT) calculations to parametrize the extended Heisenberg Hamiltonian, subsequently used in atomistic simulations to investigate the magnetic states stabilized in the three systems. Although the in-plane DMI induced in the Janus monolayers is found to be insufficient to stabilize noncollinear states, we uncover that the peculiar nature of the frustrated out-of-plane DMI favors atomic-scale $3q$ magnetic configurations stabilized at the edge of the first Brillouin zone, i.e., at very short wavelengths. 

While the stabilization of multi-$q$ magnetic states--and most commonly skyrmion lattices, has previously been reported to result from the interplay of multiple interactions, such as (frustrated) exchange and anisotropies~\cite{okubo2012multiple,lin2016ginzburg,leonov2015multiply}, dipolar interactions~\cite{yu2012magnetic}, DMI~\cite{muhlbauer2009skyrmion,yu2010real,romming2013writing}, and higher order interactions~\cite{heinze2011spontaneous,paul2020role,gutzeit2022nano,desplat2023eigenmodes}, we show here that the $3q$ state can be the ground state arising solely from out-of-plane DMI when one or two frustrated shells of nearest neighbors are considered.
We study the effect of tuning the DMI in the three systems, which can be achieved via applied strain or electric field, and report the onset of various $3q$ magnetic ground states under scaling factors as low as 3, with some configurations reminiscent of a ``nanoskyrmion"~\cite{heinze2011spontaneous} lattice.
Such atomic-scale textures, although lacking a well-defined topological charge, could exhibit peculiar transport properties, such as the nonadiatic topological Hall effect~\cite{denisov2016electron,denisov2017nontrivial}, or be useful in applications such as 2d magnonics. Our work thus adds another layer to the list of exotic spin physics already reported in FGT$_{2}$. 

\section{First principles calculations}

All first-principles calculations were performed using density functional theory (DFT) with the projector augmented wave (PAW) method \cite{PhysRevB.50.17953, PhysRevB.59.1758}, as implemented in the Vienna Ab-initio Simulation Package ({\sc vasp}) \cite{KRESSE199615, PhysRevB.54.11169}. For the exchange-correlation functional, the generalized gradient approximation (GGA) within the Perdew-Burke-Ernzerhof (PBE) formulation was used \cite{PhysRevLett.77.3865}. The plane wave basis set had a cutoff energy of 800 eV. Brillouin zone (BZ) integration was carried out using a $\Gamma$-centered Monkhorst-Pack 16 $\times$ 16 $\times$ 1 \textit{k}-point mesh \cite{PhysRevB.13.5188}. A vacuum of 30 \AA{} was introduced along the $z$-direction to prevent interactions between periodic cell images. Lattice constants and atomic positions were fully relaxed using the conjugate gradient method until the force on each atom was reduced to less than 0.001 eV \AA{}$^{-1}$. The energy convergence criterion for consecutive electronic steps was set to  $10^{-8}$ eV. 

\begin{figure}[!hbt]
    \centering\includegraphics[width=.7\columnwidth]{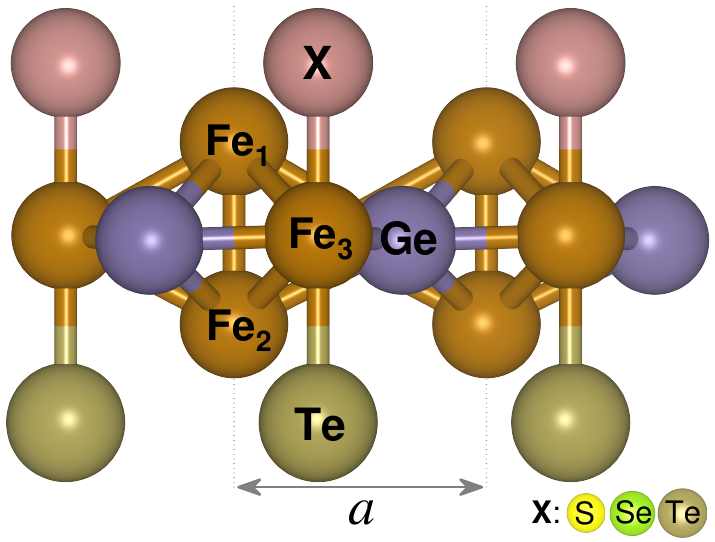} 
    \caption{Schematic representation of the atomic structure of Fe${_3}$GeXTe (X = S, Se, Te) monolayers, showing the distinct Fe, Ge, and chalcogen layers.}
    \label{fig1}
\end{figure}

The Janus FGTX (X = S, Se) monolayers (MLs) were obtained by substituting the top-layer Te atoms in pristine FGT$_2$ with S or Se. This modification breaks structural symmetry, resulting in an asymmetric atomic arrangement characteristic of Janus materials. A side view of these structures is presented in Fig.~\ref{fig1}, which illustrates the distinct layers of Fe, Ge, and chalcogen X. 

To quantify the structural effects of this asymmetry, we examined the lattice parameters of the Janus monolayers and compared them to those of pristine FGT$_2$. As summarized in Table~\ref{tab:abinito}, the lattice constants for  FGTS and FGTSe are 3.993~\AA{} and 4.023~\AA{}, respectively, both slightly smaller than those of pristine  FGT$_2$ (4.053~\AA{}). In particular, the calculated lattice parameter for  FGT$_2$ aligns with previous experimental ~\cite{deiseroth2006Fe3GeTe2,chen2013magnetic} and theoretical reports~\cite{ghosh2023unraveling,li2023tuning, kim2024strain}. To evaluate the feasibility of synthesizing Janus monolayers, we calculated their cohesive energies using the formula
    $\mathrm{E_{Coh}} = [\mathrm{E_{FGTX}}-(\mathrm{N_{Fe}}\mathrm{E_{Fe}} + \mathrm{N_{Ge}}\mathrm{E_{Ge}} + \mathrm{N_{X}}\mathrm{E_{X}} + \mathrm{N_{Te}}\mathrm{E_{Tee}})]/\mathrm{N_{atoms}}$, where $\mathrm{E_{FGTX}}$ is the total energy of the monolayer, $\mathrm{E_{Fe}}$, $\mathrm{E_{Ge}}$, $\mathrm{E_{X}}$, and $\mathrm{E_{Te}}$ are the energies of isolated Fe, Ge, X (S or Se), and Te atoms, respectively, and $\mathrm{N_{atoms}}$ represents the total number of atoms in the unit cell. As summarized in Table~\ref{tab:abinito}, the cohesive energies for FGTS and FGTSe are -3.523 eV/at. and -3.416 eV/at., respectively, both lower in energy than that of pristine FGT$_{2}$ (-3.272 eV/at.), indicating slightly enhanced bonding strength in the Janus configurations. The cohesive energies of the Janus monolayers indicate strong atomic bonding, comparable to other stable 2D materials. In line with this, we confirm the dynamical stability of the structures by calculating the phonon dispersions with detailed discussion provided in Sec.~\ref{sec:phonon} of the Supplemental Material (SM)~\cite{sm}.

    \begin{table}[!hbt]
\centering
\renewcommand{\arraystretch}{1.5} 
\setlength{\tabcolsep}{7pt}
\caption{Lattice parameters ($a$ in \AA), cohesive energies ($E_{\text{Coh}}$ in eV/atom), and intrinsic electric fields ($E_{\text{int}}$ in V/nm) for the FGTX (X = S, Se, Te) monolayers.}
\begin{ruledtabular}
\begin{tabular}{lccc} 
 & $a$ (\AA) & $E_{\text{Coh}}$ (eV/atom) & $E_{\text{int}}$ (V/nm) \\  
\hline
FGT$_2$ & 4.053  & -3.272 & -- \\
FGTSe & 4.023  & -3.416 & 1.463 \\
FGTS & 3.993  & -3.523 & 2.459 \\
\end{tabular}
\end{ruledtabular}
\label{tab:abinito}
\end{table}
    
\begin{figure}
    \centering   
    \includegraphics[width=1\linewidth]{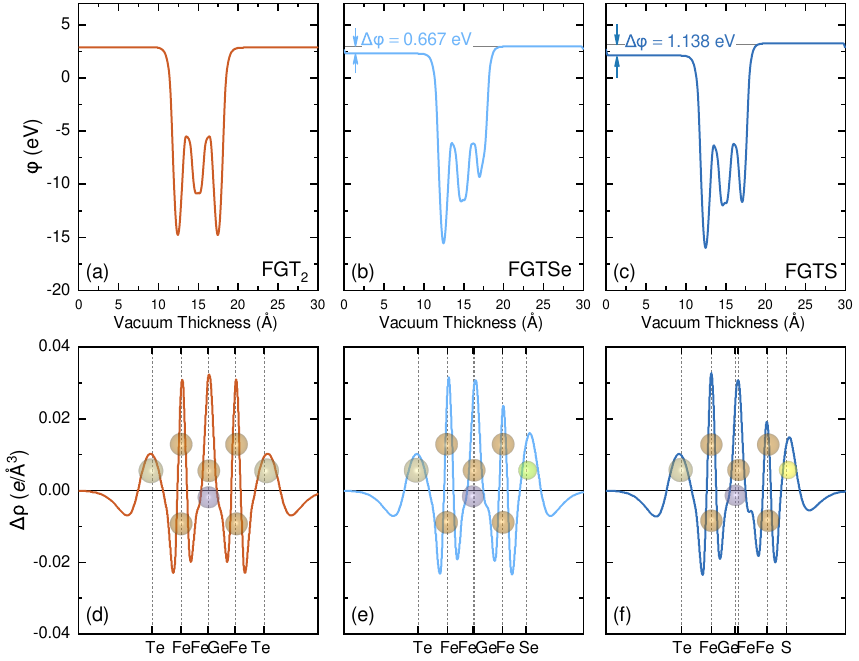} 
    \caption{(a)–(c) Electrostatic potential energy distributions for (a) FGT$_2$, (b) FGTSe, and (c) FGTS monolayers. Panels (d)–(f) show the corresponding charge density differences.}
    \label{fig:Intrinsic_Efield}
\end{figure}

Figure~\ref{fig:Intrinsic_Efield} presents the plane-averaged electrostatic potential and charge density difference (CDD) profiles for the three monolayers. The CDD is defined as $\Delta \rho ~= \rho_\mathrm{FGTX}~- \sum_i\rho^\mathrm{atom}_i$, where $\rho_\mathrm{FGTX}$ and $\rho^\mathrm{atom}_i$ are the charge densities of the monolayer and of the isolated constituent atoms, respectively. As shown in Fig.~\ref{fig:Intrinsic_Efield}(a), FGT$_2$ exhibits a symmetric electrostatic profile with no net drop across the structure. In contrast, the Janus monolayers display pronounced asymmetry arising from the broken out-of-plane mirror symmetry, as shown in Figs.~\ref{fig:Intrinsic_Efield} (b--c), with potential drops of 0.667 eV and 1.138 eV for FGTSe and FGTS, respectively, corresponding to the intrinsic electric fields ($E_{\text{int}}$) reported in Table~\ref{tab:abinito}. The CDD plots (Fig.~\ref{fig:Intrinsic_Efield}~(d--f)) show charge depletion near Fe sites and accumulation in the vicinity of the chalcogen atoms, consistent with a net charge transfer from Fe to the chalcogen atoms, confirmed by Bader charge analysis \cite{HENKELMAN2006354,https://doi.org/10.1002/jcc.20575,Tang_2009}. This charge transfer increases monotonically from  FGT$_2$ to FGTSe and FGTS, with the charge on the substituted chalcogen increasing in magnitude from -0.21~$|e|$ (Te) to -0.49~$|e|$ (Se), and -0.69~$|e|$ (S), consistent with the increasing electronegativity from Te to S (see the SM~\cite{sm} for a detailed Bader charge analysis). Correspondingly, finite intrinsic electric dipole moments of 0.053~$e$\AA{} and 0.084~$e$\AA{} are found for FGTSe and FGTS, respectively, both directed from the Te toward the substituted chalcogen surface consistent with the intrinsic electric fields given in Table~\ref{tab:abinito}. The spin-orbit coupling electronic band structures of the three monolayers are shown in Fig.~\ref{fig:socband} of the SM~\cite{sm}, where all three systems exhibit metallic behavior.

Electronic structures of FGTX monolayers, including spin-orbit coupling, were interpolated using Wannier functions obtained by projecting the electronic spectra onto Fe $3d$, Ge $4p$, Te $5p$, Se $4p$, and S $3p$ orbitals via the maximal localization technique, as implemented in the {\sc wannier90} package~\cite{MOSTOFI20142309}. Electronic states in the range from -10.0 eV to 3.5 eV relative to the Fermi level were kept frozen during the wannierization. 

The calculated values of magnetic moments are gathered in Table~\ref{table2}. For FGT$_2$, they average to 2.15~$\mu_B$/Fe. This agrees with previous first principles calculations~\cite{ghosh2023unraveling}, while being nevertheless larger than typical experimentally reported values (0.8-1.7~$\mu_B$/Fe)~\cite{chen2013magnetic,fujimura2021current,kim2022role,liu2017wafer}. This discrepancy might be due to Fe-deficiency and other structural defects present in real crystals~\cite{may2016magnetic} but absent in idealized computational models. We note that other DFT calculations using local density approximation (LDA) found 1.76~$\mu_B$/Fe, which highlights the sensitivity of the calculated moments to the exchange-correlation functional~\cite{li2023tuning}.


\section{Heisenberg Hamiltonian model}\label{sec:heisen_hamil}

To further study the properties of the three systems, we use our DFT results to parametrize the extended Heisenberg Hamiltonian,
\begin{equation}\label{eq:heisenberg_1}
\mathcal{H} = -\sum_{<ij>} J_{ij} \ {\bf S}_{i} \cdot {\bf S}_{j}  
 -\sum_{<ij>} {\bf D}_{ij} \cdot \left( {\bf S}_{i} \times {\bf S}_{j} \right) 
 - \sum_i K S_{i,z}^2,
\end{equation}
where  ${\bf S}_i$ is the normalized magnetic moment at site $i$, $J_{ij}$ is the isotropic exchange constant and ${\bf D}_{ij}$ is the Dzyaloshinskii-Moriya vector between Fe atoms $i$ and $j$, and $K$ is the uniaxial perpendicular magnetocrystalline anisotropy constant. Nearest neighbor interactions up to 21~\AA~are considered in FGT$_2$, and up to 26~\AA~in FGTSe and FGTS.

Exchange interactions were calculated based on the magnetic force theorem and the Green’s function method~\cite{LIECHTENSTEIN198765}. 
The calculations were carried out by summing over 1000 Matsubara frequencies at a temperature of 100~K~\cite{ PhysRevB.102.115162}. The calculated exchange parameters were compared to the ones obtained with TB2J~\cite{HE2021107938}, for which we found a good agreement, as shown in Fig.~\ref{fig:J_sergey_vs_tb2j} of the Supplemental Material (SM)~\cite{sm}.

The values of the main parameters, including magnetic anisotropy, leading exchange terms, and calculated Curie temperatures are summarized in Table~\ref{table2}. The exchange and anisotropy constants are given in meV/Fe atom and are found to be in general good agreement with previous first principles calculations~\cite{ghosh2023unraveling,li2023tuning}.

\begin{table*}
\centering
\caption{Parameters obtained by DFT calculations for FGT$_2$, FGTSe, and FGTS monolayers: magnetic moments $\mu_S$ in units of $\mu_{\mathrm{B}}$, perpendicular magnetocrystalline anisotropy $K$ in meV per Fe atom, first nearest-neighbor isotropic exchange interaction parameters in meV per Fe atom, and calculated Curie temperature in K, where $T_{\mathrm{C}}^\mathrm{DFT}$ and   $T_{\mathrm{C}}^\mathrm{DMFT}$ respectively denote Curie temperatures obtained using DFT, and DMFT-scaled exchange parameters.  }
\begin{ruledtabular}
\begin{tabular}{lcccccccccccc} 
 &  $\mu_{S,\mathrm{Fe_1}}$ & $\mu_{S,\mathrm{Fe_2}}$ &  $\mu_{S,\mathrm{Fe_3}}$ & $K$ & $J^1_{\mathrm{Fe_1-Fe_2}}$ & $J^1_\mathrm{Fe_2-Fe_3}$ & $J^1_{\mathrm{Fe_1-Fe_3}}$ & $J^1_{\mathrm{Fe_1-Fe_1}}$ & $J^1_{\mathrm{Fe_2-Fe_2}}$ & $J^1_{\mathrm{Fe_3-Fe_3}}$ & $T_{\mathrm{C}}^\mathrm{DFT}$ & $T_{\mathrm{C}}^\mathrm{DMFT}$  \\  
\hline
FGT$_2$ & 2.51 & 2.51 & 1.45 & 0.95 & 55.58 & 15.20 & 15.20 & -2.45 &  -2.45 &  -4.15 & 360.20 & 190.08 \\
FGTSe  &  2.71 & 2.37 &   1.33 & 0.61 & 54.92  & 9.40 & 15.23 &  -0.20 & -2.12 & -3.25 & 341.14 & 178.17 \\
FGTS &  2.83 & 2.30 & 1.24 & 0.50 & 50.25 & 6.41 & 14.86 & 0.19 & -2.51 &   -2.78 & 310.05 & 170.24 \\
\end{tabular}
\end{ruledtabular}
\label{table2}
\end{table*}


The Curie temperatures were calculated via a Monte Carlo Metropolis scheme~\cite{10.1063/1.1699114} with the Vampire atomistic framework~\cite{Evans_2014}  (see the SM for details~\cite{sm}).  For FGT$_2$, we obtain $T_\mathrm{C}^\mathrm{DFT}=360$~K, 
a value much larger than experimentally reported, where 220~K was found in bulk FGT$_2$~\cite{deiseroth2006Fe3GeTe2,chen2013magnetic}, while values between 180-200~K were measured in a few nanometer thick monolayers~\cite{fujimura2021current,guillet2024spin}. The large $T_\mathrm{C}$ obtained with our DFT parameters matches the conclusions of Ref.~\cite{ghosh2023unraveling}, where it was shown that dynamical mean field theory (DMFT) yields more physical results than DFT for the FGT$_{2}$ family. The isotropic exchange and DMI constants from DFT were reported to be overestimated but remain qualitatively the same as the DMFT values, with global scaling factors between them of  $J_\mathrm{DMFT}\approx  0.54 J_\mathrm{DFT}$ for exchange, and $D_\mathrm{DMFT}\approx 0.70 D_\mathrm{DFT}$ for DMI. This observation motivates the following treatment: in the rest of this work, we consider both the DFT parameters, and these same parameters scaled by the above factors, which we refer to as ``DMFT scaled". Since the DMFT exchange couplings are approximately 2 times smaller than the DFT ones, the Curie temperatures are reduced by a similar factor and are closer to experimental values for FGT$_2$, for which we find $T_C^\mathrm{DMFT}=190~K$~\cite{sm}. Note that in Ref.~\cite{ghosh2023unraveling}, the authors report for FGT$_2$ a magnetocrystalline anisotropy of $K=1.5$~meV/Fe with DFT, and $K=1.25$~meV/Fe with DMFT. As this is a small variation compared to that of exchange, and in general the contribution of anisotropy to the total energy of the systems is small, we did not apply scaling on the anisotropy, and simply used the DFT values throughout the manuscript.

In what follows, we discuss the exchange and DMI parameters obtained from DFT in FGT$_2$, and how they are modified in the Janus MLs.

\subsection{Isotropic Exchange}
\begin{figure}
\includegraphics[width=1\linewidth]{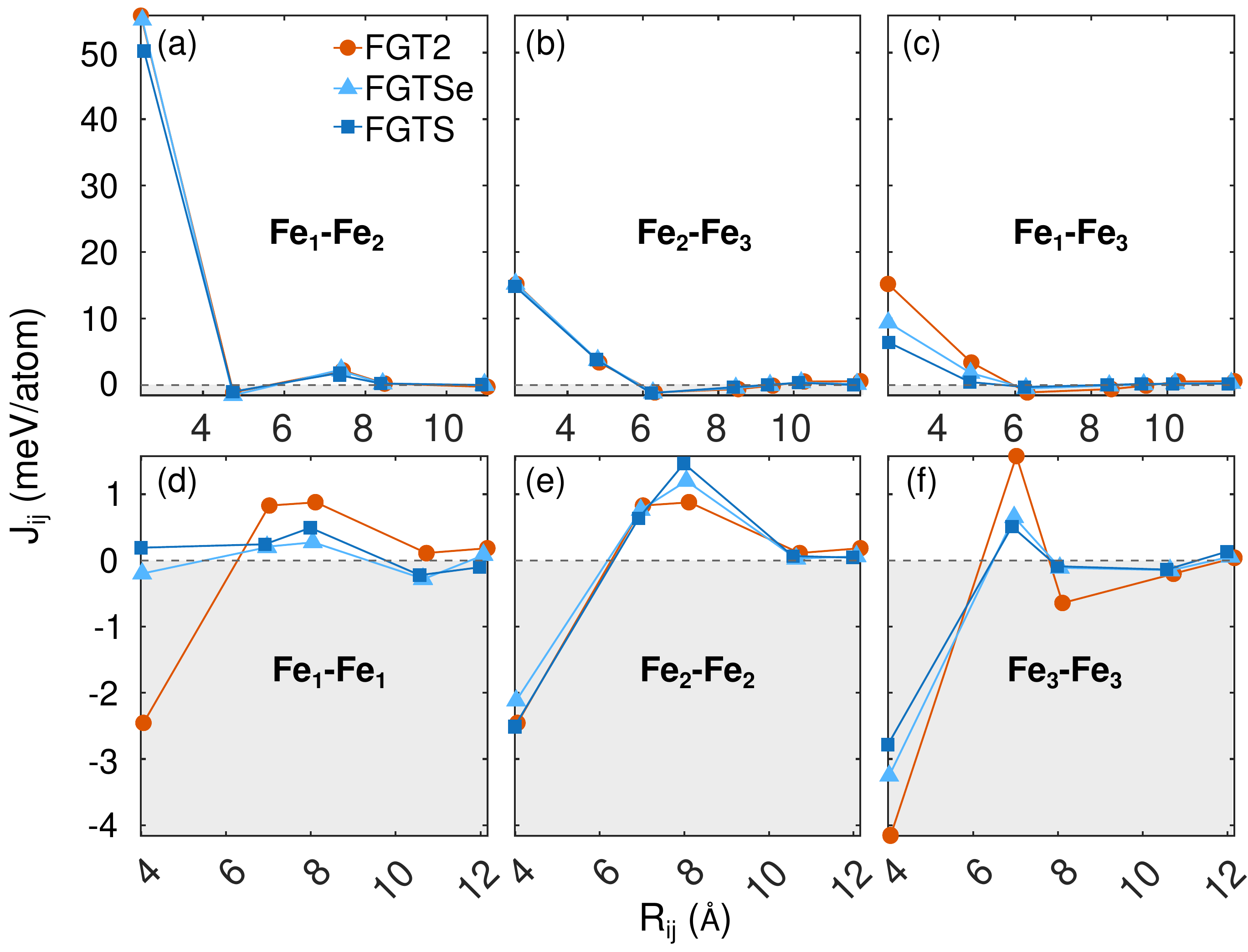}
\caption{Isotropic exchange constant as a function of neighbor distance $R_{ij}$ in the three systems up to a distance of 12~\AA, where (a--c) correspond to interlayer exchange, and  (d--f) correspond to intralayer exchange. Exchange is given in meV per Fe atom, where $J_{ij}>0 (<0)$ indicates (anti)ferromagnetic coupling.\label{fig:exchange}}  
\end{figure}

In addition to the first nearest-neighbor site interactions values gathered in Table~\ref{table2}, the isotropic exchange interactions for nearest-neighbor distances up to 12~\AA~ are shown in Fig.~\ref{fig:exchange} for all three systems. We adopt the convention where $J>0 (<0)$ indicates (anti)ferromagnetic coupling.

As reported previously~\cite{li2023tuning,ghosh2023unraveling}, the interlayer first nearest neighbor coupling,  $J^1_{\mathrm{Fe_1-Fe_2}}$, where the superscript $^1$ indicates first nearest neighbor, is found to be the largest exchange value (50-55~meV/at) in the three systems (Fig.~\ref{fig:exchange}(a)).  Meanwhile, the intralayer exchange at the first nearest neighbor level is antiferromagnetic (Fig.~\ref{fig:exchange}(d--f)). The exchange amplitude then decreases with distance, while exhibiting sign oscillations indicative of exchange frustration. Overall, we find that the introduction of S and Se atoms does not seem to impact the exchange coupling amplitude significantly. 

\subsection{DMI}

\begin{figure*}
\includegraphics[width=.7\linewidth]{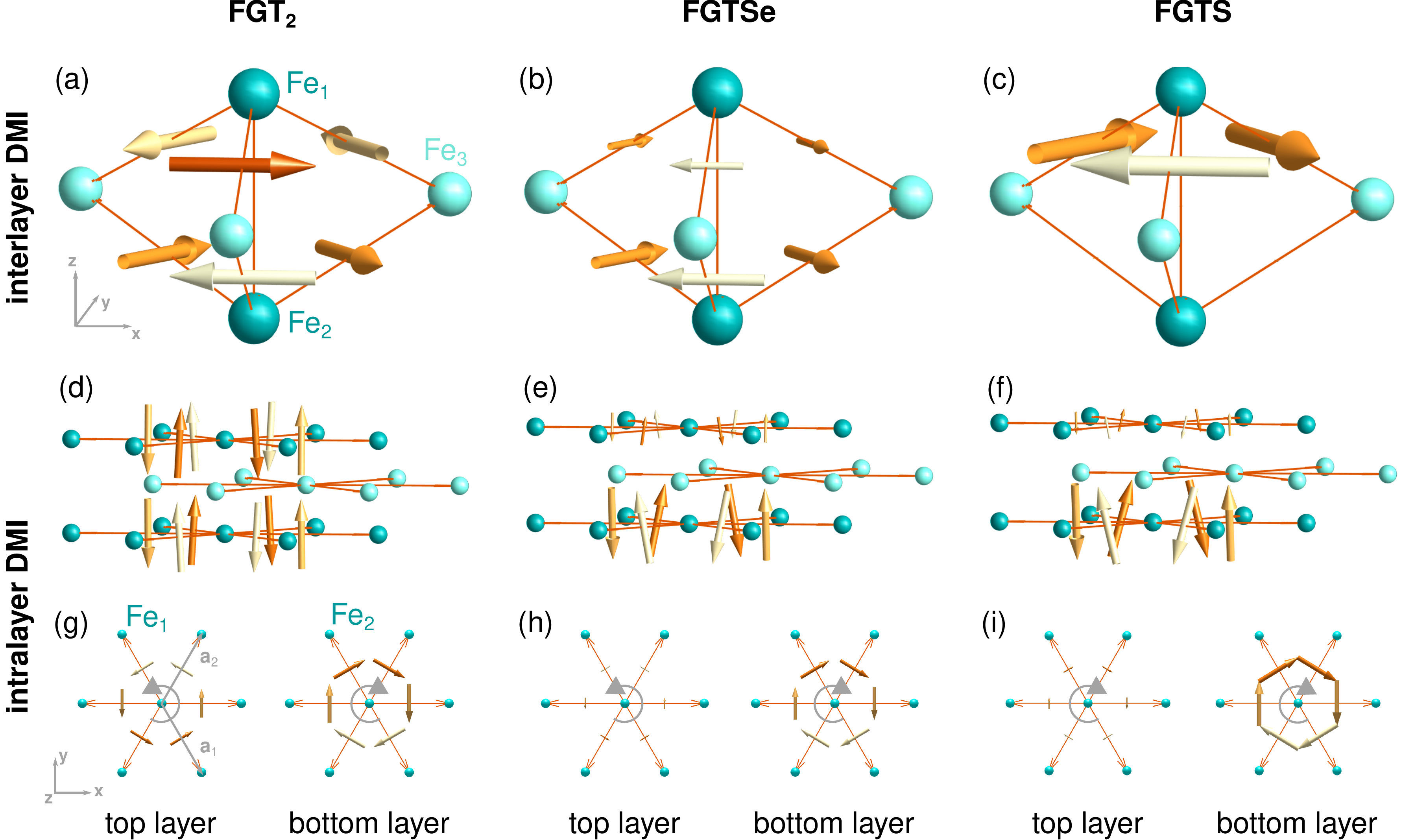}
\caption{Anatomy of the DMI in (a, d, g) FGT$_2$, (b, e, h) FGTSe, and (c, f, i) FGTS. (a--c) show the interlayer DMI in the first unit cell, while (d--i) show the intralayer DMI between neighboring sites separated by one lattice constant. All components of the DMI vectors are plotted in (a--f), while in (g--i), purely the in-plane components are shown, and scaled up for readability. The DMI vectors are colored according to their polar angle in the XY plane. The solid orange arrows indicate the direction of the bonds, and point away from the central atom. The solid grey semicircles in (g--i) indicates the in-plane chirality of the DM vectors. Note that for readability, the scaling applied to the DMI vectors is consistent within each subfigure, but not between different subfigures. In (g), we also show the in-plane Bravais vectors in real space, $\hat{\bf a}_1$ and $\hat{\bf a}_2$.
\label{fig:dmi_sketch}}  
\end{figure*}

\begin{figure}
    \includegraphics[width=1\linewidth]{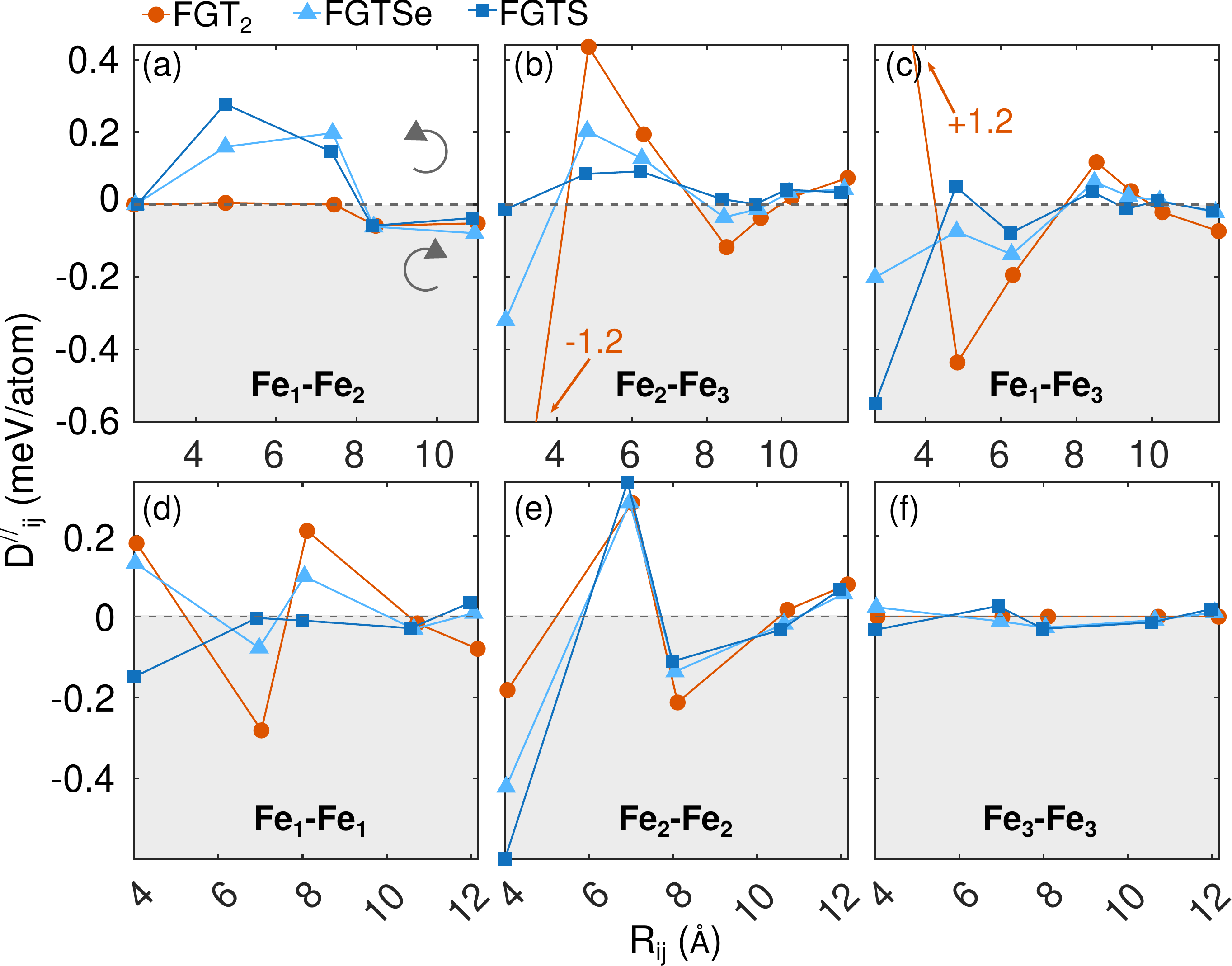} 
\caption{Amplitude of the in-plane DMI components as a function of neighbor distance  in the three systems up to a distance of 12~\AA, where (a--c) correspond to interlayer DMI, and  (d--f) correspond to intralayer DMI. The DMI amplitude is given is meV per Fe atom, where $D^\parallel_{ij}>0 (<0)$ indicates counterclockwise (clockwise) chirality of the in-plane DM vectors.\label{fig:dmi_ip}}  
\end{figure}

\begin{figure}
    \includegraphics[width=1\linewidth]{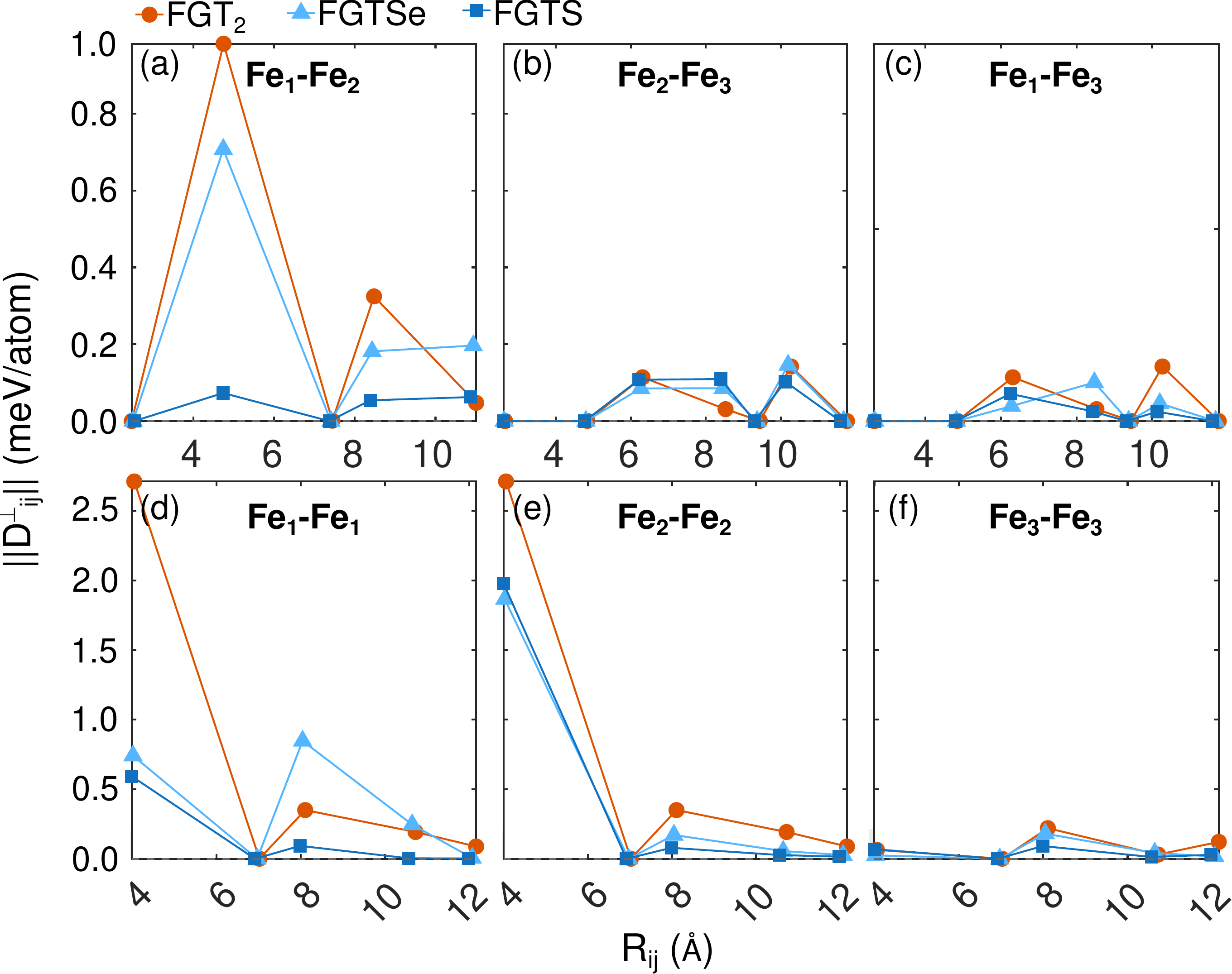} 
    \caption{Absolute value of the out-of-plane DMI components as a function of neighbor distance given is in meV per Fe atom in the three systems up to a distance of 12~\AA, where (a--c) correspond to interlayer DMI, and  (d--f) correspond to intralayer DMI. Note that within a single a shell of neighbors, the sign of the OOP components alternates depending on bond direction, so there is no defined chirality. \label{fig:dmi_oop} }  
\end{figure}

In Fig.~\ref{fig:dmi_sketch}, we sketch the direction of the DM vectors for FGT$_2$ (Figs.~\ref{fig:dmi_sketch}(a, d, g)), FGTSe (Figs.~\ref{fig:dmi_sketch}(b, e, h)), and FGTS (Figs.~\ref{fig:dmi_sketch}(c, f, i)) for first interlayer (a--c) and intralayer (d--i) neighbors. 
The in-plane amplitude of the DM vectors, defined as $D^\parallel_{ij}=\sqrt{D_{ij,x}^2+D_{ij,y}^2}$, is plotted in Fig.~\ref{fig:dmi_ip}, 
with the sign corresponding to the chirality of the DM vectors, whereby $D^\parallel_{ij}>0 (<0)$ indicates counterclockwise (clockwise) chirality. 
Similarly, we show the absolute value of the out-of-plane (OOP) components of the DMI in Fig.~\ref{fig:dmi_oop}. Note that within a single shell of neighbors, the sign of the OOP components alternates depending on bond direction, as visible for instance in Figs.~\ref{fig:dmi_sketch}(d--f), so there is no defined chirality.

%

\paragraph{First neighbors interlayer:} In all systems, the  interlayer DMI within the unit cell is purely in plane, with the geometry of interfacial DMI that favors  N\'eel type spin textures rotating out of plane, i.e., ${\bf D}_{ij}=D^\parallel_{ij} \left( {\bf \hat{R}}^\parallel_{ij} \times {\bf \hat{z}} \right)$, with ${\bf \hat{R}}_{ij}^\parallel$ the in-plane component of the unit vector in the direction of the bond between sites $i$ and $j$, and ${\bf \hat{z}}$ the out-of-plane direction (Figs.~\ref{fig:dmi_sketch}(a--c)). 
In FGT$_2$,  $D^1_{\mathrm{Fe_1-Fe_3}}$ and  $D^1_{\mathrm{Fe_3-Fe_2}}$, have the same amplitude with opposite chirality due to the (001) mirror plane (see Figs.~\ref{fig:dmi_ip}(b,c)), and $D^1_{\mathrm{Fe_1-Fe_2}}\approx0$,  resulting in zero net DMI within the unit cell (uc)~\cite{laref2020elusive}.
This is not the case in the Janus MLs, as in FGTSe,  $D^1_{\mathrm{Fe_1-Fe_3}}$ and  $D^1_{\mathrm{Fe_2-Fe_3}}$ exhibit the same chirality, with an amplitude of -0.3 and -0.2~meV/at.  (Figs.~\ref{fig:dmi_ip}(b,c)), while in FGTS, $D^1_{\mathrm{Fe_3-Fe_2}}$ is negligible compared to $D^1_{\mathrm{Fe_1-Fe_3}}=-0.5$~meV/at, resulting in both cases in nonzero effective in-plane DMI within the uc.

\paragraph{First neighbors intralayer:}Contrary to the interlayer DMI, the intralayer DMI between first nearest neighbors in all three systems is mainly found out of plane (Figs.~\ref{fig:dmi_sketch}(d--f)). 
In general, the DMI in the middle layer, $D_{\mathrm{Fe_3-Fe_3}}$, remains negligible compared to that of the top and bottom layers (Figs.~\ref{fig:dmi_oop}(d--f)).  We find the largest OOP component for FGT$_2$, with   $\lvert \lvert D^1_{\mathrm{Fe_1-Fe_1}} \rvert \rvert =\lvert \lvert D^1_{\mathrm{Fe_2-Fe_2}} \rvert \rvert =2.7$~meV/at. This is reduced in the Janus MLs, with a more pronounced effect on the top Fe$_1$ layer that interfaces with S or Se. As mentioned above, the DM vectors alternatively point along $+z$ and $-z$ depending on the direction of the bond, which results in a form of frustration. 

The in-plane components of the intralayer DMI are sketched in Figs.~\ref{fig:dmi_sketch}(g--i), with the corresponding amplitudes plotted in Fig.~\ref{fig:dmi_ip}(d--f). Similar to the interlayer DMI, they cancel out in FGT$_2$ between the top and bottom layers, but remain finite in the Janus MLs. However, their values are small compared to the OOP components, with the largest amplitude found for $D^{1,\parallel}_{\mathrm{Fe_2-Fe_2}}=-0.6$~meV/at in FGTS, which also has the same chirality as, and therefore adds up to, $D^{1,\parallel}_{\mathrm{Fe_1-Fe_1}}=-0.2$~meV/at.

The rest of the components then decrease in amplitude with nearest neighbor distance with some sign oscillations, yielding an additional form of DMI frustration.

While it has been reported that the out-of-plane  DMI does not play a significant role on the stabilization of skyrmions in 2D magnets~\cite{liang2020very,du2022spontaneous}, it should in principle favor N\'eel-type modulated states rotating in the plane~\cite{laref2020elusive}. This will be investigated in what follows.

\section{Dispersion of $1q$ spin spirals}

\begin{figure*}
\includegraphics[width=.7\linewidth]{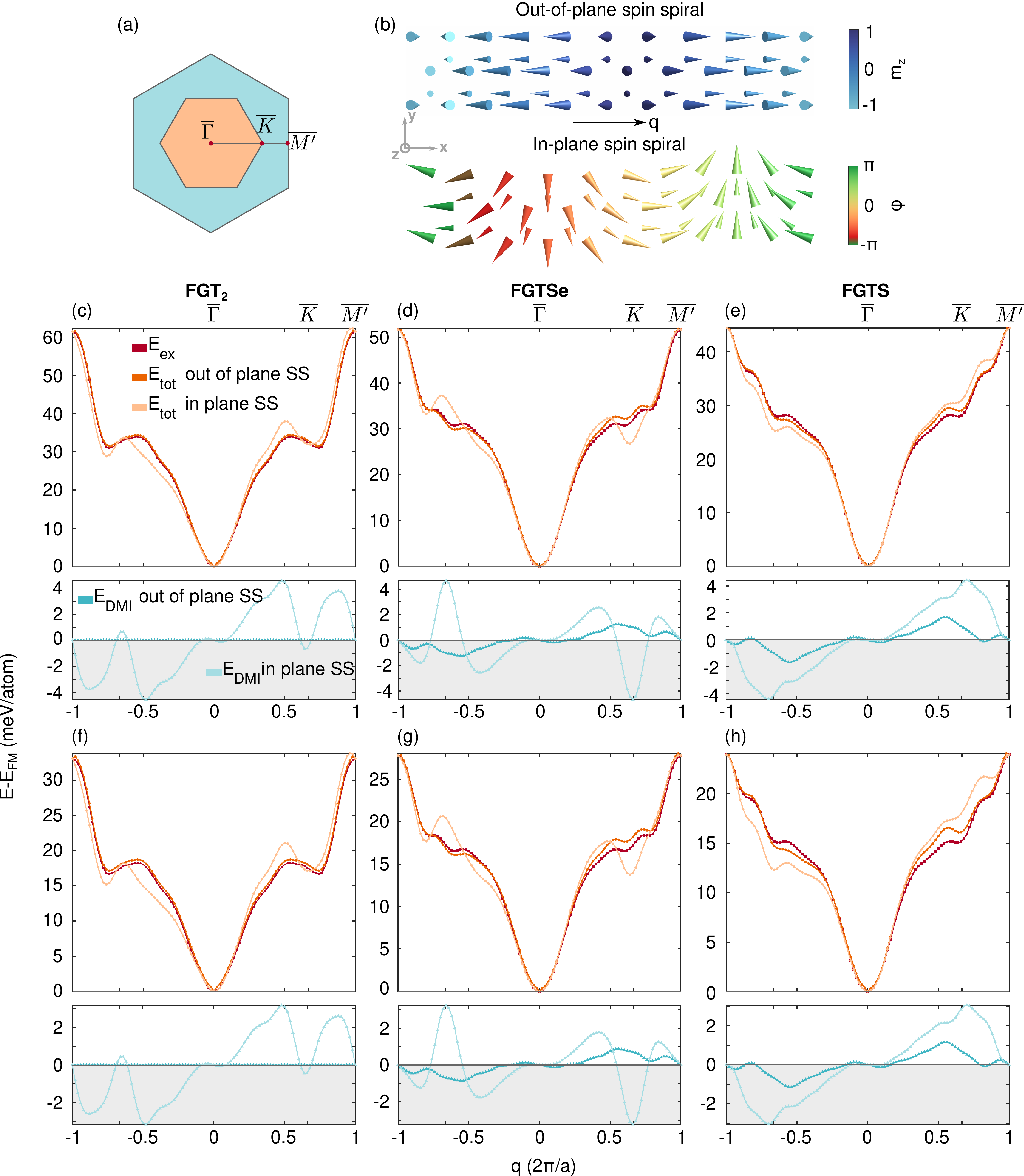}
\caption{Energy dispersion of N\'eel-type spin spirals propagating along the $\overline{ \Gamma KM'}$ high symmetry direction. (a) Sketch of the first (orange) and second (blue) 2D Brillouin zones, where the considered high-symmetry points are marked. (b) Spin configuration corresponding to N\'eel spin spirals rotating out of plane (top) and in plane (bottom), shown here for $q=0.1$. The spins carried by Fe$_3$ atoms are shown in a smaller size.  (c--h) Energy of spin spirals relative to that of the ferromagnetic state $E_{\mathrm{FM}}$, for (c, f) FGT$_2$, (d, g) FGTSe, (e, h) FGTS, where we  used (c--e) DFT parameters, (f--h) DMFT-scaled parameters ~\cite{ghosh2023unraveling}. For each subfigure, we show at the top the total energy, $E_{\mathrm{tot}}$, of out-of-plane (dark orange) and in-plane (light orange) spirals, and the isotropic exchange energy, $E_{\mathrm{ex}}$ (red). On the bottom graphs, we show the DMI energy of out-of-plane (dark blue) and in-plane (light blue) spirals, $E_{\mathrm{DMI}}$. Exchange and DMI are considered up to 21~\AA~in FGT$_2$, and up to 26~\AA~in FGTSe and FGTS.\label{fig:dispersions}}  
\end{figure*}

We use the Spirit atomistic framework~\cite{muller2019spirit} to compute the energy of N\'eel-type spin spirals propagating along the $\overline{\Gamma K}$  high-symmetry direction. The presence of 3 Fe atoms in the unit cell allows for the existence of states at shorter wavelengths than the edge of the first Brillouin zone, and so we compute the dispersions up to $\overline{M}'$ in the second BZ, as sketched in Fig.~\ref{fig:dispersions}(a).

The magnetization at real space position ${\bf r}$ is given by ${\bf S} ({\bf r})= {\bf R}_q \cos ({\bf q \cdot r}) + {\bf I}_q \sin ({\bf q \cdot r})$, with ${\bf R}_q={\bf \hat{x}} \parallel {\bf q} $, where ${\bf \hat{x}}$ is the Cartesian unit vector (Fig.~\ref{fig:dispersions}(b)).  The in-plane DMI components favor N\'eel spirals rotating out of plane, obtained by setting ${\bf I}_q={\bf \hat{z}}$ (Fig.~\ref{fig:dispersions}(b), top), while the out-of-plane components favor spirals rotating in the plane, which are obtained by setting  ${\bf I}_q={\bf \hat{y}}$  (Fig.~\ref{fig:dispersions}(b), bottom). We thus consider the energy of both types of spin spirals. In the rest of this work, the magnetic supercell consists of 100 $\times$ 100 unit cells, i.e., 300 $\times$ 300 Fe atoms. 

The obtained dispersions are shown in Figs.~\ref{fig:dispersions}(c, f) for FGT$_2$, \ref{fig:dispersions}(d, g) for FGTSe, and \ref{fig:dispersions}(e, h) for FGTS, where Figs.~\ref{fig:dispersions}(c--e) correspond to DFT parameters, and Figs.~\ref{fig:dispersions}(f--h) correspond to scaled parameters according to DMFT~\cite{ghosh2023unraveling}. 
In all cases, the total energy dispersions exhibit a global minimum at $\overline{\Gamma}$  ($q=0$), corresponding to the ferromagnetic state. Due to the steep dispersion of isotropic exchange near $\overline{\Gamma}$, and the fact that the DMI is quenched around $\overline{\Gamma}$~\cite{laref2020elusive} and exhibits global minima at the edge of the first BZ, or even in the second BZ, we find that noncollinear states are not stabilized with the current parameters. In the following, we look in further detail at the DMI dispersions.

\subsection{Out-of-plane spirals}
We first analyze the DMI dispersion of out-of-plane spirals (dark blue curves in Figs~\ref{fig:dispersions}(c--h)), which is only sensitive to the in-plane DMI components. As expected from the (001) mirror plane, the DMI energy is zero in FGT$_2$, while it assumes nonzero values in FGTSe and FGTS.
Due to frustration in the in-plane components of the DM vectors (Fig.~\ref{fig:dmi_ip}), the energy dispersion does not resemble the typical sine curve yielded by first-nearest-neighbor N\'eel-type DMI, with extrema at the center of the 1st BZ. 
Instead, we find a local shallow minimum for right-rotating  spin spirals at $q=0.1$ in FGTSe, and $q=0.08$ in FGTS, and a global minimum for left-rotating spin spirals at $q=-0.56$ in FGTSe with energy $E^\mathrm{min}_{\mathrm{DMI}}=-1.25$~meV/at (-0.86~meV/at for DMFT-scaled DMI), and at $q=-0.54$ in FGTS, with $E^\mathrm{min}_{\mathrm{DMI}}=-1.67$~meV/at (-1.15~meV/at for DMFT-scaled DMI). In the SM~\cite{sm}, we show the shell-resolved DMI dispersion of out-of-plane spirals, detailing how DMI frustration yields the present dispersion curves.

\subsection{In-plane spirals}

Next, the dispersion of in-plane spin spirals, sensitive to the out-of-plane DMI components, is shown in light blue in Figs.~\ref{fig:dispersions}(c--h). Unlike its in-plane counterpart, the out-of-plane DMI does not cancel out in FGT$_2$. It possesses multiple local minima, 
and a global one at  $q=-0.48$, with energy $E^\mathrm{min}_{\mathrm{DMI}}=-4.57$ meV/at ($-3.15$ meV/at for DMFT-scaled DMI). A similar case is found for FGTSe, with global minima coinciding with the $\overline{K}$ point, at $q=0.66$, with $E^\mathrm{min}_{\mathrm{DMI}}=-4.7$~meV/at ($-3.22$~meV/at for DMFT). In FGTS, the global minimum is found in the 2nd BZ, at $q=-0.7$, with $E^\mathrm{min}_{\mathrm{DMI}}=-4.42$~meV/at ($-3.05$~meV/at for DMFT).

\begin{figure}
\includegraphics[width=1\linewidth]{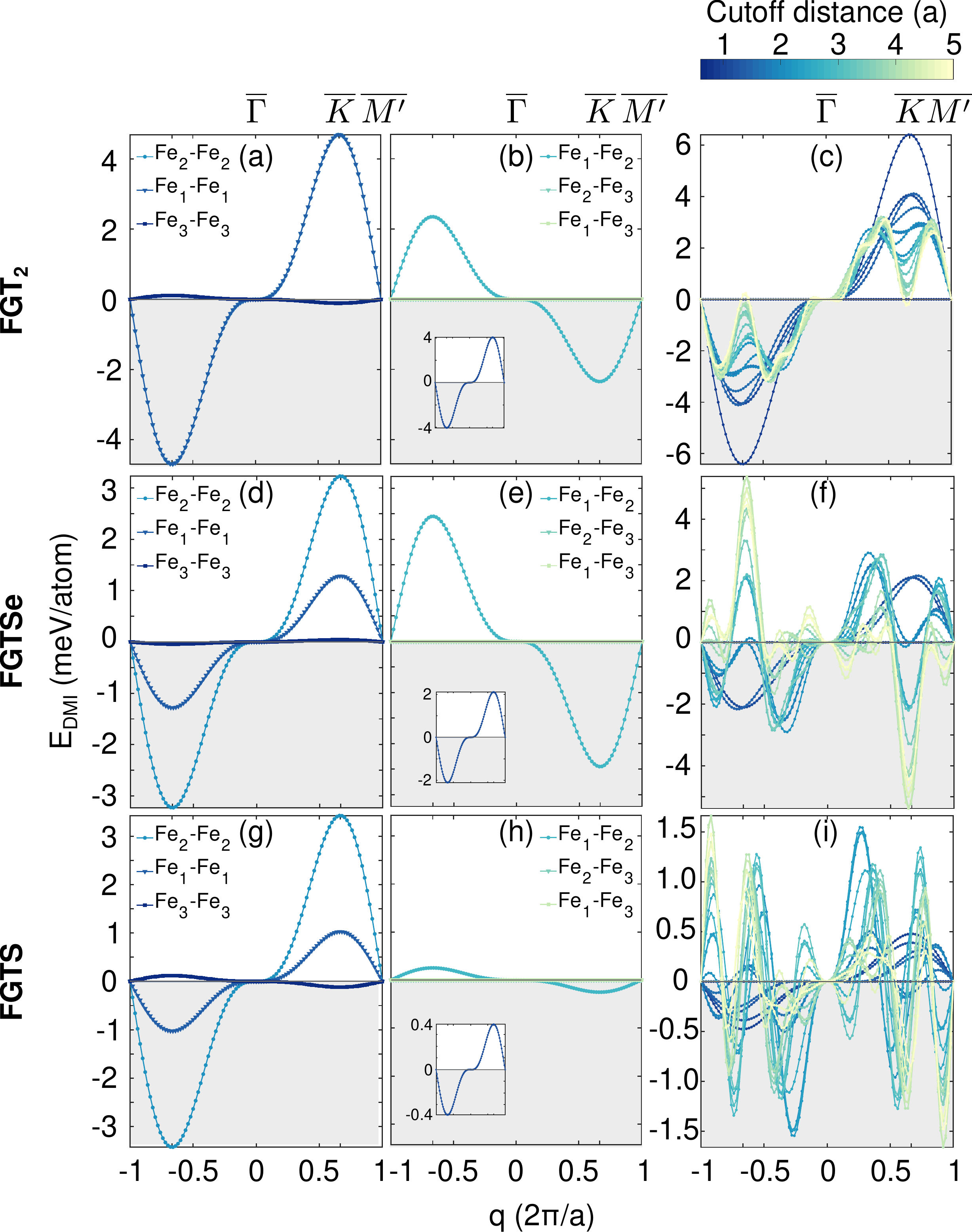}
\caption{Shell-resolved DMI Energy of N\'eel-type out-of-plane spin spirals propagating along the $\overline{ \Gamma KM'}$  high symmetry direction as sketched in Fig.~\ref{fig:dispersions}(a) for (a--c) FGT$_2$, (d--f) FGTSe, (g--j) FGTS for DFT (unscaled) DMI parameters. We show contributions of (a, b, g) intralayer first nearest neighbors and (b, e, h) interlayer second nearest neighbors, where the insets in (b, e,  h) show the sum of intra- and interlayer contributions. (c, f, i) Total DMI energy dispersion as a function of cutoff distance, expressed in units of the lattice constant $a$.  Note that we only show the dispersion up to a distance of $5a$, whereas the dispersions in Fig.~\ref{fig:dispersions} consider neighbor interactions beyond $6a$.
\label{fig:Edmi_dispersion_shells_axisz} 
}
\end{figure}

In order to explain the shape of the DMI dispersion for in-plane spirals, we show in Fig.~\ref{fig:Edmi_dispersion_shells_axisz} the contributions to the DMI energy of spin spirals from the first intralayer 
(Figs.~\ref{fig:Edmi_dispersion_shells_axisz}(a, d, g)) and second interlayer  nearest neighbors (Figs.~\ref{fig:Edmi_dispersion_shells_axisz}(b, e, h)), in each case respectively for FGT$_2$, FGTSe and FGTS.
Note that the DMI components between interlayer  first nearest neighbors are purely in-plane (Figs.~\ref{fig:dmi_oop}(a--c)), and thus do not contribute to the energy of in-plane rotating spin spirals. 

For the first neighbors within the layers, the OOP DMI components along ${\bf \hat{a}}_{1,2}$ bonds have opposite signs to the ones along the ${\bf \hat{a}}_1+{\bf \hat{a}}_2$ bonds (Figs~\ref{fig:dmi_sketch}(d--f)). The DMI energy dispersion along $\overline{\Gamma K}$ then assumes extrema at $q=\pm 2/3$,  which coincides with the $\overline{K}$ point. 
Over the extended 2D BZ, we can derive its analytical form as, 
\begin{equation}\label{eq:2d_oop_dmi_disp}
    \begin{split}
    E^{\perp}_{\mathrm{DMI}} (q_x,q_y)=& -2D^\perp_{{\bf \hat{a}}_1+{\bf \hat{a}}_2} 
    (  \sin(\pi (q_x+\sqrt{3}q_y) + \\
    & \sin(\pi (q_x-\sqrt{3}q_y) - 
    \sin(2 \pi q_x) ),
    \end{split}
\end{equation}
where $D^\perp_{{\bf \hat{a}}_1+{\bf \hat{a}}_2}$ denotes the amplitude and sign of the OOP DMI component along the ${\bf \hat{a}}_1 +{\bf \hat{a}}_2$ bond. 

Eq.~(\ref{eq:2d_oop_dmi_disp}) is plotted in Fig.~\ref{fig:dmi_states_shellresolved}(a) over the extended BZ. It possesses three minima and three maxima that coincide with the $\overline{K}$ points of the first BZ, where the nature of the points (minimum or maximum) is determined by the sign of $D^\perp_{{\bf \hat{a}}_1+{\bf \hat{a}}_2}$. This implies that atomic-scale spatially modulated states propagating along three $\overline{\Gamma K}$ directions are equally favored by the DMI. Between the lobes, the inflection points where the DMI energy vanishes are found along the $\overline{\Gamma M}$ direction, which is a direction of mirror symmetry~\cite{laref2020elusive}.  Note that we find from the DFT that this is not strictly the case when all shells of neighbors are considered, but the DMI amplitude along $\overline{\Gamma M}$ remains over an order of magnitude smaller than along $\overline{\Gamma K}$.

For the second-neighbor interlayer interactions, we find the only nonzero OOP component 
for the Fe$_1$-Fe$_2$ pairs  (Figs.~\ref{fig:dmi_oop}(a--c)),  so the dispersion assumes the same profile shown in Fig.~\ref{fig:dmi_states_shellresolved}(a).

Last, in Figs.~\ref{fig:Edmi_dispersion_shells_axisz}(c, f, i), we show the cumulative DMI dispersions as a function of cutoff distance, up to 5 lattice constants $a$. This illustrates how the complex profile of the global DMI dispersions shown in Figs.~\ref{fig:dispersions} emerges from the contributions of many inter- and intralayer neighbors at different distances, resulting in a sum of sines with different periodicities. The sum of all contributions also provides deeper or shallower minima compared to the contributions from the first few shells of neighbors (insets in Figs.~\ref{fig:Edmi_dispersion_shells_axisz}(b, e, h)), showcasing the the importance of considering more distant neighbors.

In the next section, we examine the nature of the magnetic textures favored by the DMI in these systems.


\section{Emergence of a $3q$ DMI ground state}\label{sec:dmi_31_ground_state}

 In the following, we study how the ground state of the DMI evolves in the three systems, when different shells of nearest neighbors are taken into account. To do so, the exchange and anisotropy are set to 0, and states are relaxed with the velocity projection solver~\cite{bessarab2015method,muller2019spirit} starting from the corresponding minimum single-$q$ configuration according to Fig.~\ref{fig:Edmi_dispersion_shells_axisz}. 
 
 \subsection{FGT$_2$} 
\begin{figure}
\includegraphics[width=1\linewidth]{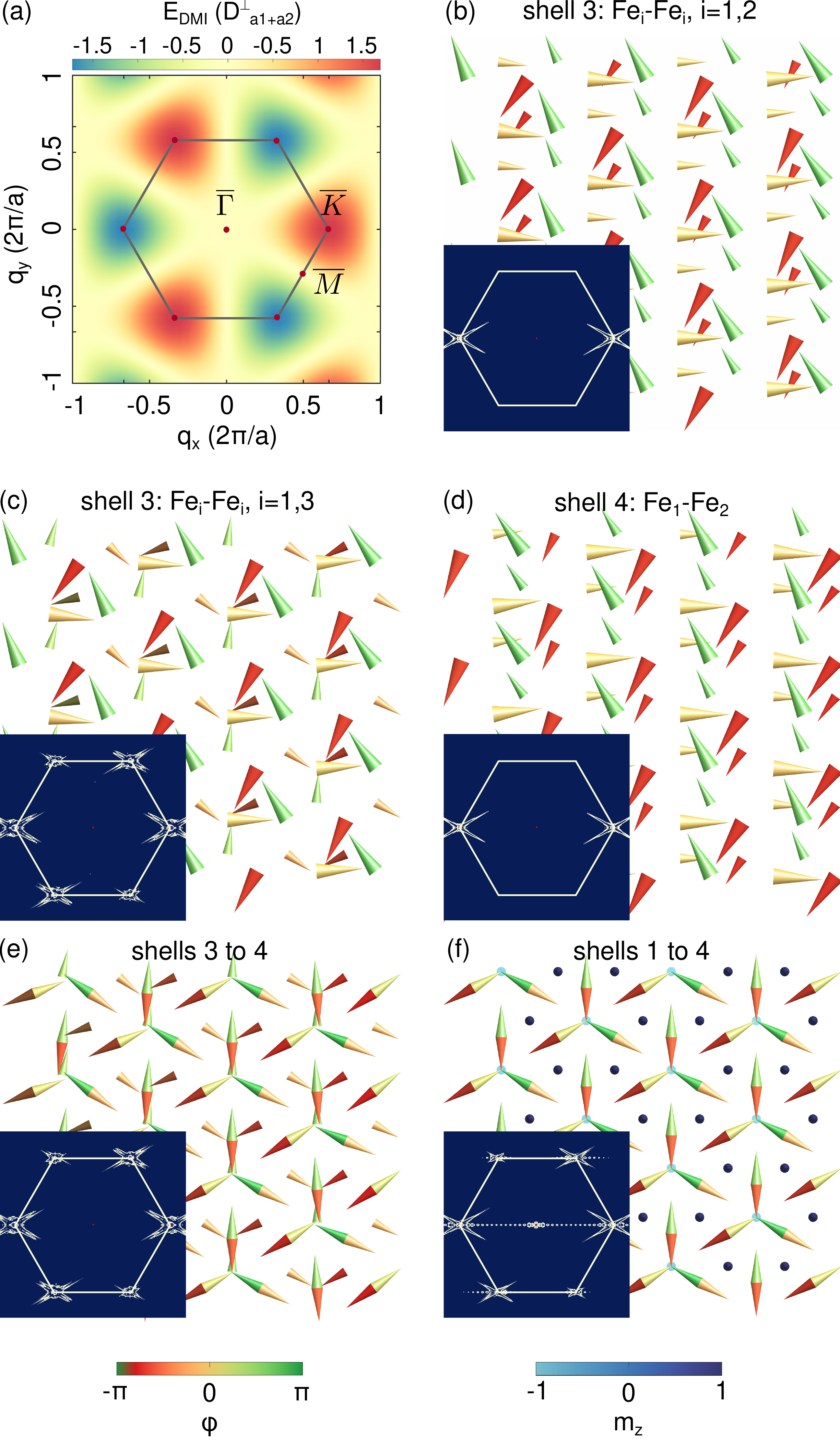}
\caption{In FGT$_2$ (a) profile of the DMI Energy from shells 3 to 4 of $1q$ spin spirals rotating in plane over the extended 2D BZ according to Eq.~(\ref{eq:2d_oop_dmi_disp}), given in units of the perpendicular DMI component along the ${\bf \hat{a}}_1+{\bf \hat{a}}_2$ bond, $D^\perp_{{\bf \hat{a}}_1+{\bf \hat{a}}_2}$. (b--f) Zoomed-in spin configuration corresponding to the DMI ground state  for (b, c) shell 3 corresponding to first neighbors intralayer Fe$_{i}$-Fe$_{i}$ where (b) $i=1,2$, (c) $i=1,3$, (d) shell 4 corresponding to second neighbors interlayer, Fe$_{1}$-Fe$_{2}$ (e) shells 3 and 4, (f) shells 1 to 4. The exchange and anisotropy are set to 0. Spins carried by Fe$_3$ atoms are shown in a smaller size. The insets in (b--f) show the extended Brillouin zone, where the peaks correspond to the amplitudes resulting from the spatial FFT of the magnetic texture. Note that half of the peaks are complex conjugates. 
\label{fig:dmi_states_shellresolved} 
}
\end{figure}

We first focus on FGT$_2$, for which a zoomed-in portion of the resulting spin configurations is plotted in Figs.~\ref{fig:dmi_states_shellresolved}(b--f). For each case, we show the corresponding 2D fast Fourier transform  (FFT) of the magnetization profile, with the 1st BZ superimposed as a guide to the eye. Note that the $z$-axis atomic coordinates are not taken into account as we construct the extended 2D BZ.

  In Fig.~\ref{fig:dmi_states_shellresolved}(b), we show the spin configuration corresponding to the single-$q$, in-plane spiral propagating along $\overline{\Gamma K}$ with $q=-2/3$, which is the ground state of the DM interaction when only $D^1_\mathrm{Fe_i-Fe_i}$, where $i=1,2$, are considered (Fig.~\ref{fig:Edmi_dispersion_shells_axisz}(a)).

In contrast, Fig.~\ref{fig:dmi_states_shellresolved}(c) shows the state relaxed when all first intralayer neibghbors in shell 3 are included (Fig.~\ref{fig:Edmi_dispersion_shells_axisz}(a)), i.e, $D^1_\mathrm{Fe_i-Fe_i}$, where $i=1,3$. Since $D^1_\mathrm{Fe_{1,2}-Fe_{1,2}}$ and $D^1_\mathrm{Fe_{3}-Fe_{3}}$ have opposite signs, in the absence of interlayer couplings, single-$q$ states of opposite $q$ vectors are relaxed in the different layers. As shown by the FFT in the inset, the resulting texture is in fact a planar $3q$ state, where the $q$ vectors are found along three $\overline{\Gamma K}$ directions of the 1st BZ forming ${120}^{\circ}$ angles with equal amplitudes $q_i=2/3$, for $i=1,3$. The competing nature of the DMI between different sub-lattices does not transpire from the single-$q$ dispersions, from which one would simply assume a single-$q$ energy minimum. However, the 3-fold symmetry of the DMI dispersion over the extended 2D BZ (Fig.~\ref{fig:dmi_states_shellresolved}(a)) confirms that three $\overline{\Gamma K}$ directions are favorable.

  Fig.~\ref{fig:dmi_states_shellresolved}(d) shows the ground state for shell 4, corresponding to second neighbors intralayer, i.e., $D^2_\mathrm{Fe_{1}-Fe_{2}}$. In agreement with the dispersion in Fig.~\ref{fig:Edmi_dispersion_shells_axisz}(b), it is the single-$q$ state along $\overline{ \Gamma{K}}$ with $q=+2/3$.

Next, Fig.~\ref{fig:dmi_states_shellresolved}(e) shows the ground state of the DMI from shells 3 and 4, yielding a different planar $3q$ state. The spins carried by Fe$_1$ and Fe$_2$ atoms in the same unit cell are antiparallel to each other. This allows simultaneous clockwise spin rotation ($q>0$) between the Fe$_1$ and Fe$_2$ atoms, and the counterclockwise rotation ($q<0$) within each individual Fe$_1$ and Fe$_2$ layer. Note that in such a case where Fe$_1$ and Fe$_2$ spins are antiparallel with the same periodicity, we restrict the FFT to Fe$_1$ and Fe$_3$ sublattices only. 

Last, Fig.~\ref{fig:dmi_states_shellresolved}(f) shows the state obtained by additionally accounting for the first interlayer nearest neighbors, i.e, $D^1_{\mathrm{Fe}_2-\mathrm{Fe}_3}$ and  $D^1_{\mathrm{Fe}_1-\mathrm{Fe}_3}$ (given that $D^1_{\mathrm{Fe}_1-\mathrm{Fe}_2}$ vanishes), corresponding to shells 1 to 4. Since these additional DM vectors lie in-plane (Fig.~\ref{fig:dmi_sketch}(a)), the OOP-rotating spirals are favored between Fe$_{1,2}$ and Fe$_3$ spins. As a consequence, Fe$_3$ spins point out of plane with alternating signs, and the resulting $3q$ state is no longer purely planar.

\subsection{Janus monolayers } 
\begin{figure}
\includegraphics[width=1\linewidth]{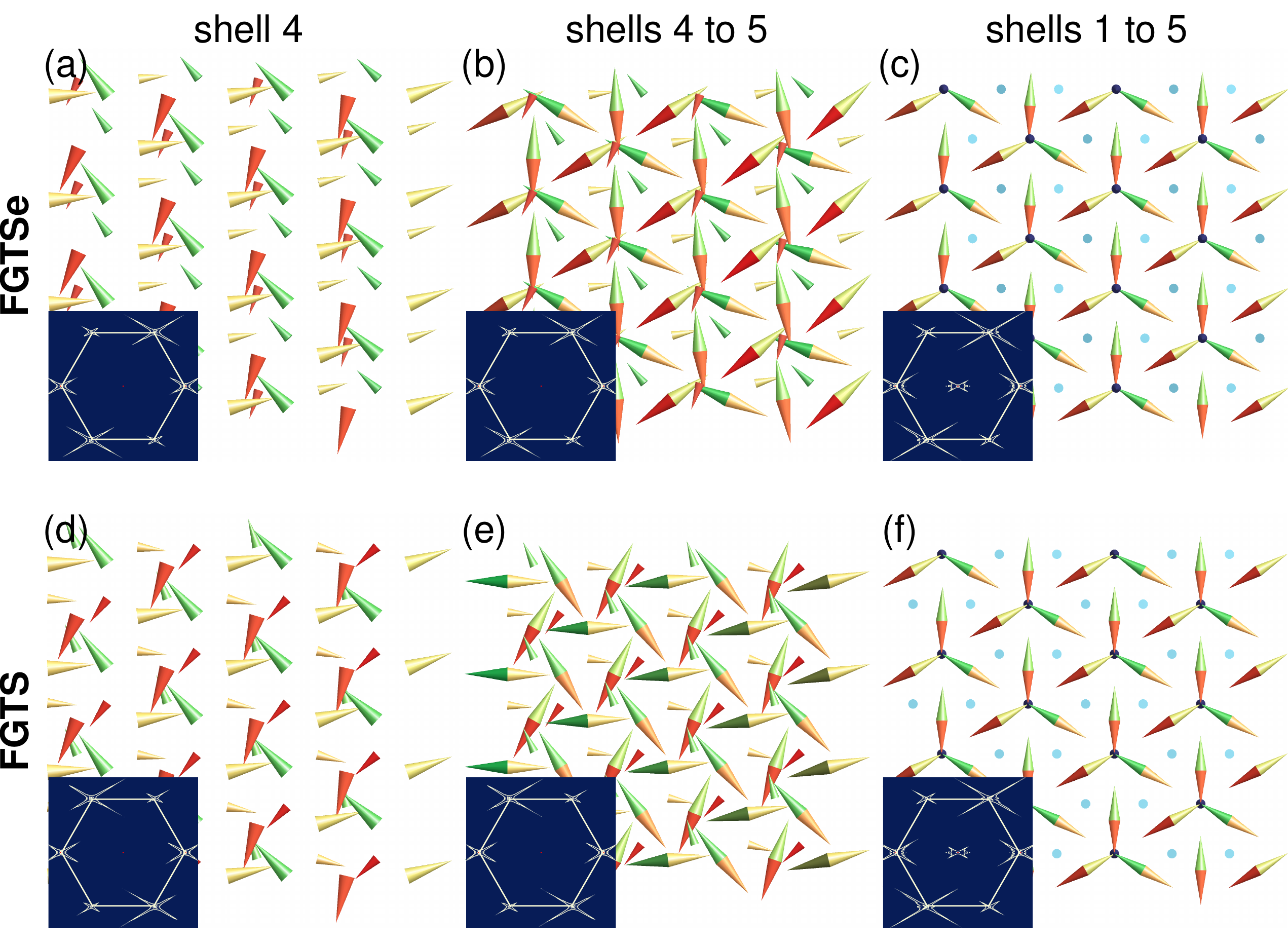}
\caption{Zoomed-in spin configuration corresponding to the DMI ground state in (a--c) FGTSe, and (d--f) FGTS for (a, d) shell 4 corresponding to first neighbors intralayer Fe$_{i}$-Fe$_{i}$ where $i=1,3$, (b, e) shell 4 to 5, where shell 5 corresponds to second neighbors interlayer, Fe$_{1}$-Fe$_{2}$, (c, f) shells 1 to 5.  The exchange and anisotropy are set to 0. Spins carried by Fe$_3$ atoms are shown in a smaller size. The insets show the extended Brillouin zone, where the peaks correspond to the amplitudes resulting from the spatial FFT of the magnetic texture.
\label{fig:janus_dmi_states_shellresolved} 
}
\end{figure}

Similar states are obtained in the Janus MLs, which we gather in Fig.~\ref{fig:janus_dmi_states_shellresolved}.
Note that the broken out-of-plane symmetry results in asymmetric nearest neighbor distances between the top and bottom layers, so the shell numbering is shifted compared to FGT$_2$. Similar to FGT$_2$, we obtain the planar $3q$ states from the first intralayer neighbors (shell 4, Figs.~\ref{fig:janus_dmi_states_shellresolved}(a, d)), the planar $3q$ states with antiparallel Fe$_1$ and Fe$_2$ spins by including the second interlayer neighbors  (shells 4 to 5, Figs.~\ref{fig:janus_dmi_states_shellresolved}(b, e)), and the non-planar $3q$ states with shells 1 to 5 (Figs.~\ref{fig:janus_dmi_states_shellresolved}(c, f)).

Unlike in FGT$_2$,  we find that the $q$-vector amplitudes deviate slightly from 2/3. The magnetization acquires a weak OOP component due to the uncompensated in-plane DMI. The wavevector imposed by the latter, $q\approx 0.3$ (see the SM~\cite{sm}), most likely interferes with that of the OOP components. The resulting period is no longer exactly one lattice constant, leading to spatially varying magnetic textures across neighboring unit cells, and states not typically commensurate with the magnetic supercell. We only show a zoomed-in version chosen over an arbitrary region.

\subsection{``Nanoskyrmion" lattice}
The relaxed  non-planar $3q$ states such as shown in Figs.~\ref{fig:dmi_states_shellresolved}(f)  and \ref{fig:janus_dmi_states_shellresolved}(c, f)  resemble  (antiferro)magnetic  skyrmion lattices (skX) at the nanoscale, although topological charge cannot be defined properly with the 90-degree angles between neighboring spins~\cite{berg1981definition}. The ferromagnetic counterpart to this texture can be analytically described as a sum of three spirals~\cite{okubo2012multiple},
\begin{equation}\label{eq:skx}
\begin{split}
    S_x & =I_x \sum_{j=1}^3  \sin ({\bf q}_j \cdot {\bf r} + \theta_j )\hat{e}_{x}^j,\\
    S_y & =I_y \sum_{j=1}^3 \sin ({\bf q}_j \cdot {\bf r} + \theta_j )\hat{e}_{y}^j,\\
    S_z & =I_z \sum_{j=1}^3 \cos ({\bf q}_j \cdot {\bf r} + \theta_j ),
\end{split}
\end{equation}
where $j=1,2,3$ runs over the 3 $q$-vectors, $I_{x,y,z}$ are normalizing constants,  $\theta_j$ are phase factors, and ${\bf \hat{e}}_j$ are normalized in-plane vectors. Here, the $q$ vectors are along the real space Bravais vectors with 120 degrees between them, ${\bf q}_1=q(-1/2,\sqrt{3}/2,0)  \parallel -{\bf \hat{a}}_1$,  ${\bf q}_2 =q(-1/2,-\sqrt{3}/2,0) \parallel -{\bf \hat{a}}_2$, and ${\bf q}_3 =q(1,0,0)\parallel ({\bf \hat{a}}_1+{\bf \hat{a}}_2)$, with $q=\pm2/3$. The  ${\bf \hat{e}}_j$ vectors correspond to a N\'eel configuration with ${\bf \hat{e}}_j \parallel {\bf q}_j$. The  $\theta_j$ allow to shift the out-of-plane spins over to the Fe$_3$ sublattice with $\left(\theta_1,\theta_2,\theta_3 \right)=-\pi q (0,-1,1)$. For smaller amplitudes of the wavevector closer to $\overline{\Gamma}$, Eq.~(\ref{eq:skx}) generates a regular skyrmion lattice.

\subsection{DMI energy}
In Table~\ref{tab:E_dmi}, we gather the DMI energy of the $3q$ states plotted in Figs.~\ref{fig:dmi_states_shellresolved} and ~\ref{fig:janus_dmi_states_shellresolved} using the DMFT-scaled parameters, and compare them with that of the $1q$ state that minimizes the DMI energy 
(Fig~\ref{fig:Edmi_dispersion_shells_axisz}).
\begin{table}
\caption{In FGT$_2$, FGTSe, and FGTS with DMFT-scaled parameters, DMI energy in meV/at. of the lowest energy $1q$ state compared to that of the relaxed $3q$ DMI ground state when different shells of nearest neighbors are taken into account. The corresponding figure showing the relaxed $3q$ spin configuration is given as indication for each case. \label{tab:E_dmi}}
\begin{ruledtabular}
\begin{tabular}{l|l|ccc}
   & Shell(s) & 3 (\ref{fig:dmi_states_shellresolved}c)  & 3 to 4 (\ref{fig:dmi_states_shellresolved}e)   & 1 to 4  (\ref{fig:dmi_states_shellresolved}f) \\  
    \cline{2-5}
\multirow{3}{3em}{FGT$_2$} & $E_{\mathrm{DMI}}$($1q$) &  
-6.41  & -4.06  & -4.06 \\
& $E_{\mathrm{DMI}}$($3q$) &    -6.56 &  -8.91 &  -11.01 \\
\colrule
&  Shell(s)  & 4  (\ref{fig:janus_dmi_states_shellresolved}a)  & 4 to 5 (\ref{fig:janus_dmi_states_shellresolved}b)   & 1 to 5  (\ref{fig:janus_dmi_states_shellresolved}c) \\  
 \cline{2-5}
 \multirow{3}{3em}{FGTSe} & $E_{\mathrm{DMI}}$($1q$) &   -3.09 &  -1.43 &   -1.43 \\
& $E_{\mathrm{DMI}}$($3q$) &  -3.14 &   -4.83 &   -4.92 \\
\colrule
&  Shell(s) & 4  (\ref{fig:janus_dmi_states_shellresolved}d)  & 4 to 5  (\ref{fig:janus_dmi_states_shellresolved}e)   & 1 to 5   (\ref{fig:janus_dmi_states_shellresolved}f) \\  
 \cline{2-5}
\multirow{3}{3em}{FGTS} & $E_{\mathrm{DMI}}$($1q$) & -2.94   & -2.77 &   -2.77 \\
& $E_{\mathrm{DMI}}$($3q$) & -3.16 &   -3.34 &   -3.74  \\
\end{tabular}
\end{ruledtabular}
\end{table}
For reference, Table~\ref{tab:E_dmi_dft} in the SM~\cite{sm} contains the same data for unscaled DFT parameters.

In FGT$_2$, the DMI energy of the relaxed $3q$ state when shells 1 to 4 are taken into account is around -11~meV/at, which represents a reduction in DMI energy of approximately 5~meV/at. from the lowest energy $1q$ state. 
Similarly, in the Janus MLs, the $3q$ states represent an energy reduction of about 4 and 1~meV/at., respectively in FGTSe and FGTS, compared to the $1q$ states.
Compared to the exchange energy at the edge of the 1st BZ in the 3 systems, $E_\mathrm{ex} \approx 17$~meV/at (Figs.\ref{fig:dispersions}(f--h)), one may infer that scaling  the DMI by about 3 in FGT$_2$, and about 5 in the Janus MLs, may stabilize noncollinear states in these systems.

The effect of scaling the DMI components is formally explored in the next section.

\section{Tuning the DMI}\label{sec:scalingD}

The possibility of tuning the DMI was predicted in 2D materials via strain~\cite{cui2020strain,shen2022strain} or applied electric field~\cite{behera2019magnetic,liu2018electrical,liu2018analysis}. For instance, 8\% biaxial strain was theoretically predicted to increase the DMI by over a factor of 2 in Janus Cr$_2$I$_3$Y$_3$ (Y=Br, Cl)~\cite{shen2022strain}. Similarly, in FGT/Ge heterostructures, ab initio calculations predicted that a small compressive strain of 3\% can enhance  the microscopic (in-plane) DMI by 400\%~\cite{li2023tuning}. In CrI$_3$, a perpendicular electric field was shown to break the inversion symmetry and induce in-plane DMI, whereby a field of 2~V/nm was predicted to result in a DMI of -0.8~meV~\cite{liu2018analysis}. 
The in-plane mirror symmetry in FGTX monolayers could be further broken by an in-plane electric field~\cite{wang2019giant} or applied uniaxial strain~\cite{niu2024reducing}, allowing control over the out-of-plane DMI.


\subsection{$1q$ dispersion}
\begin{figure}
\includegraphics[width=1\linewidth]{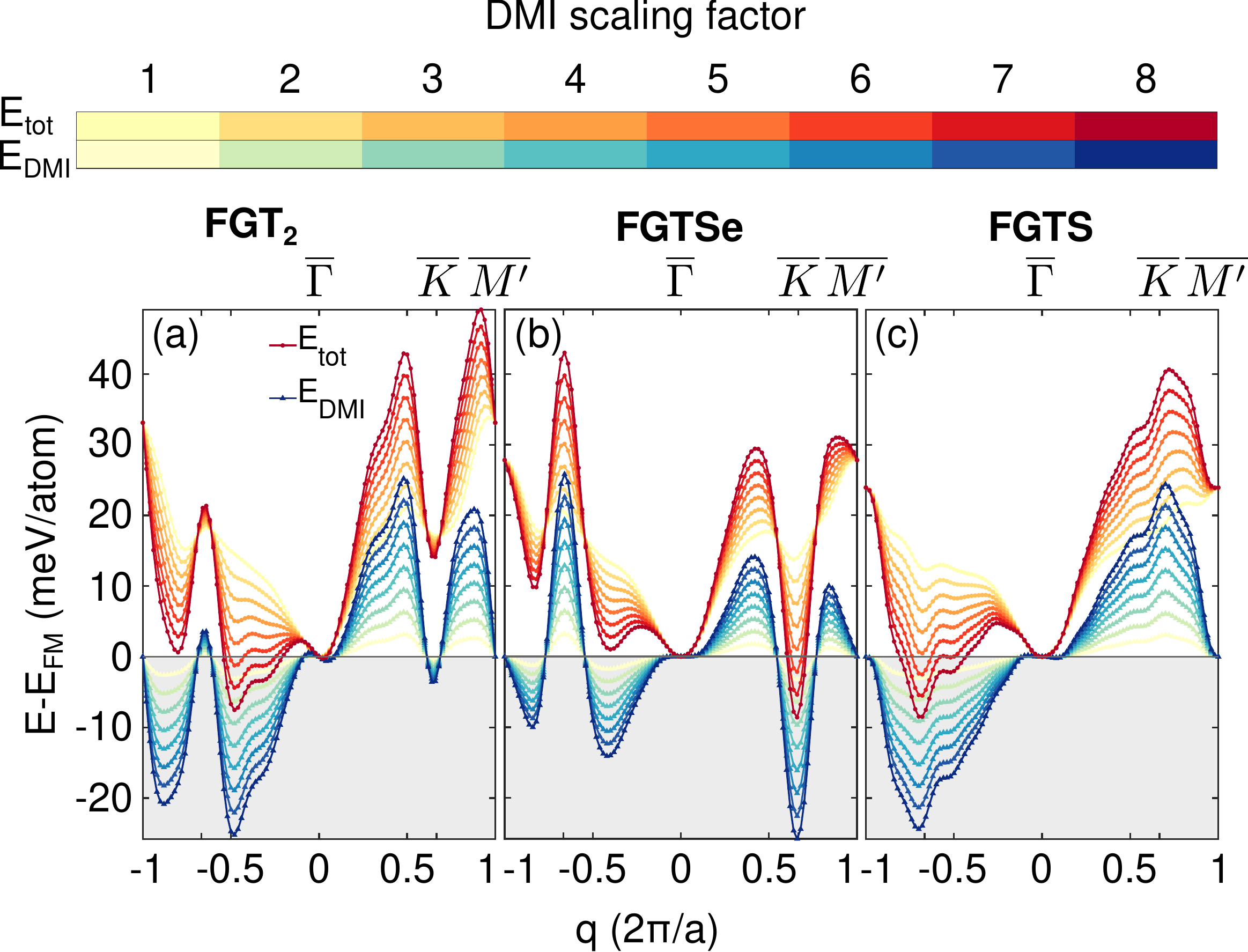}
\caption{Effect of scaling the amplitude of the DMI on the energy of single-$q$ N\'eel spin spirals rotating in-plane relative to the ferromagnetic state, $E_\mathrm{FM}$, in (a) FGT$_2$ (b) FGTSe, (c) FGTS. The scaled parameters according to DMFT are used. We show the total energy, $E_{\mathrm{tot}}$, and the DMI energy, $E_{\mathrm{DMI}}$. Their values upon scaling the DMI are given by the colormaps at the top, where the yellow to red colormap corresponds to $E_{\mathrm{tot}}$, and the yellow to blue  colormap corresponds to $E_{\mathrm{DMI}}$. 
\label{fig:dispersions_scaleD_axisz}
}  
\end{figure}

In this section, exchange and anisotropy are reincorporated in the Hamiltonian. In the three systems, we multiply all in-plane and out-of-plane components of the DMI by a common scaling factor between 1 and 8 and compute the resulting dispersion of single-$q$ N\'eel spin spirals propagating in-plane. 
The results are shown in Fig.~\ref{fig:dispersions_scaleD_axisz} using DMFT-scaled parameters. Unscaled parameters lead to similar conclusions.
In FGT$_2$ (Fig.~\ref{fig:dispersions_scaleD_axisz}(a)), scaling the DMI by a factor of 3 and up to 5 yields a minimum close to $\overline{\Gamma}$ at $q=0.02$, while a factor of 6 and above yields a minimum at $q=-0.48$. In the Janus MLs (Figs.~\ref{fig:dispersions_scaleD_axisz}(b, c)), we find global minima at the edge of the 1st BZ, at $q=0.66$ in FGTSe, and $q=-0.68$ in FGTS, for DMI scaling of 6 and above.

 In the SM~\cite{sm}, we similarly show the dispersion of  out-of-plane-rotating spin spirals, for both scaled and unscaled parameters in the Janus MLs. Since the in-plane components have much smaller amplitudes, much larger, a priori experimentally unrealistic scaling factors between 14 to 19 are necessary to obtain energy minima away from $\overline{\Gamma}$. 

\subsection{$3q$ ground states}

\begin{figure}
\includegraphics[width=1\linewidth]{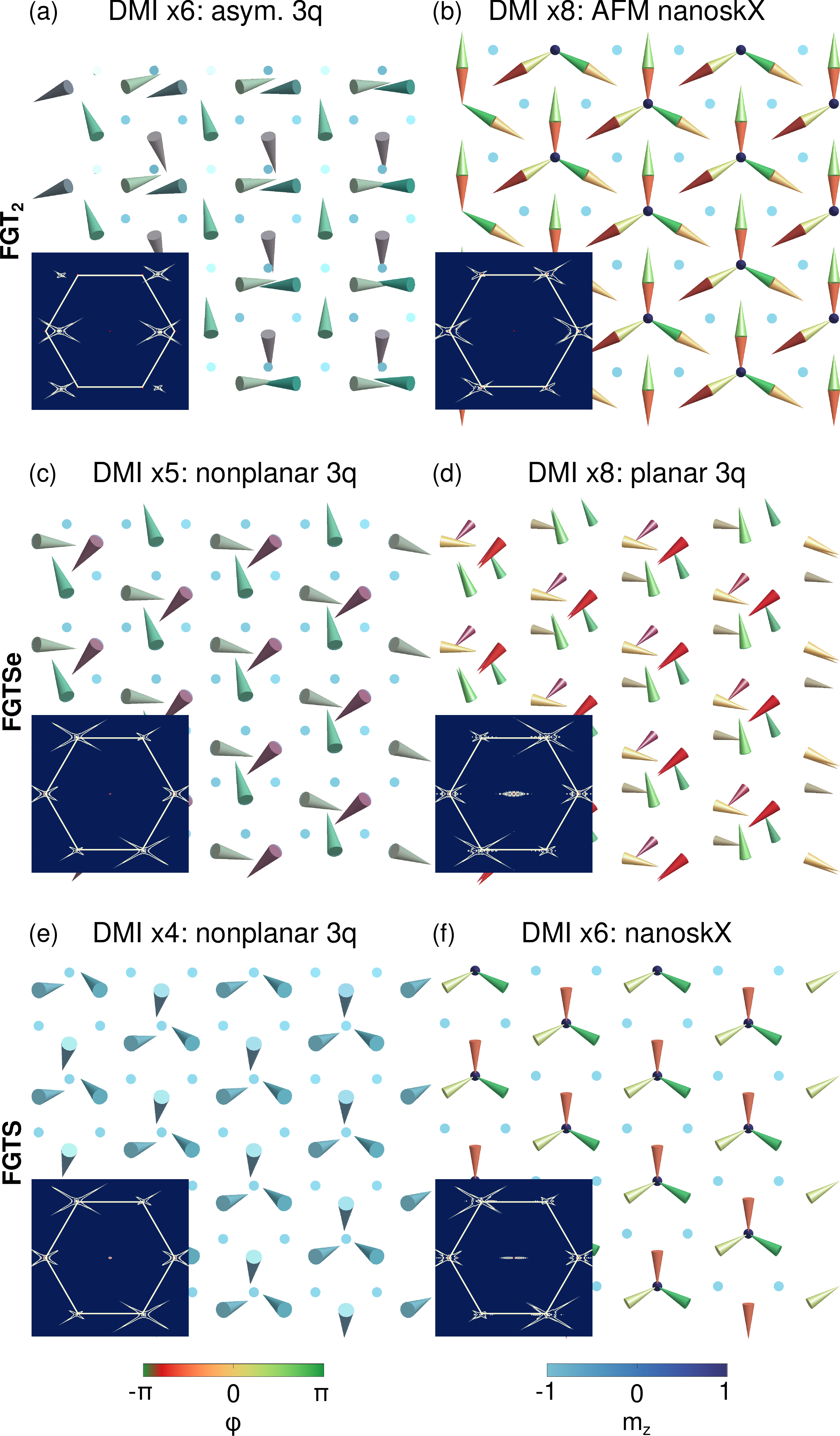}
\caption{Zoomed-in portion of the magnetic states relaxed with DMFT-scaled parameters in (a, b) FGT$_2$ (c,d) FGTSe, (e, f) FGTS for different values of the DMI scaling factor. Each subplot is an example of the phases described in Fig.~\ref{fig:energy_contrib_Dscaling}. Spins carried by Fe$_3$ atoms are plotted with a smaller size.  The insets show the extended Brillouin zone, where the peaks correspond to the amplitudes resulting from the spatial FFT of the magnetic texture.
\label{fig:3q_spins}  }
\end{figure}

As demonstrated in the previous section, the frustrated out-of-plane DMI tends to favor $3q$ states, implying that the dispersion of single-$q$ spin spirals is not a good metric to probe the states of the considered systems. We therefore iteratively increase the DMI scaling factor and relax the magnetic configuration with the velocity projection solver, starting from the ferromagnetic state at a scaling factor of 1. Examples of the relaxed  states are shown in Fig.~\ref{fig:3q_spins}.

\begin{figure}
\includegraphics[width=1\linewidth]{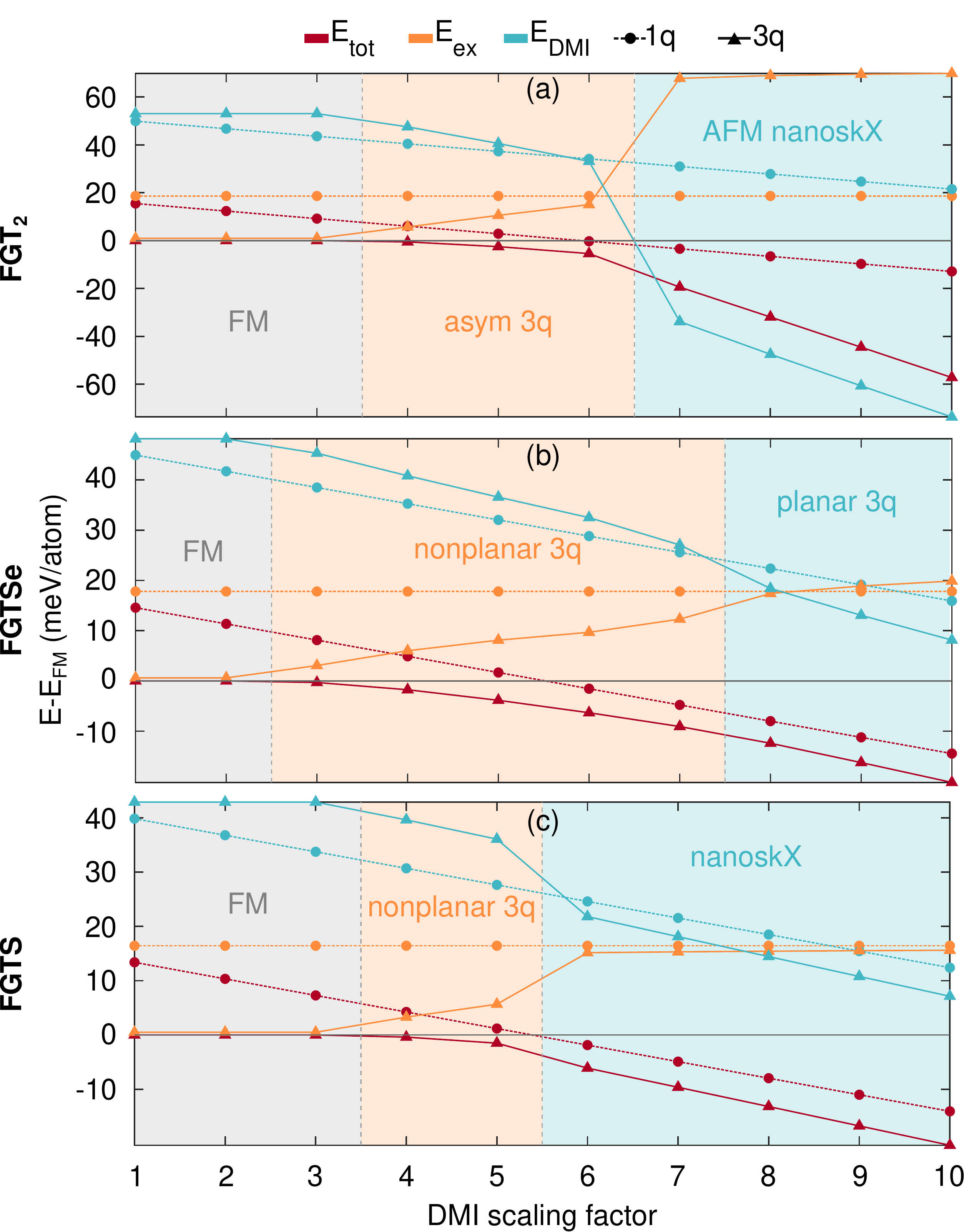}
\caption{Energy contributions to $1q$ and $3q$ states relative to the ferromagnetic state, $E_\mathrm{FM}$, as a function of the DMI scaling factor with DMFT-scaled parameters in (a) FGT$_2$, (b) FGTSe, (c) FGTS. For both states, we show the total energy ($E_\mathrm{tot}$, red markers), the exchange energy ($E_\mathrm{ex}$, orange markers), and the DMI energy ($E_\mathrm{DMI}$, blue markers). The $1q$ state is chosen as the minimum in the DMI energy dispersion (Fig.~\ref{fig:dispersions}), while the $3q$ state results from energy minimization.
\label{fig:energy_contrib_Dscaling}  }
\end{figure}

In Fig.~\ref{fig:energy_contrib_Dscaling}, we give the total energy and individual DMI and exchange contributions of the relaxed states  as a function of the DMI scaling factor, and we compare them with that of the $1q$ state that minimizes the DMI dispersion energy (Figs.~\ref{fig:dispersions}(f--h)).  We focus on DMFT-scaled parameters.
Note that the anisotropy is of the order of 1~meV/at. in all three systems, and thus does not contribute to the following discussion. 

Indeed, we find that the $1q$ state is never the ground state of the total energy.  At low DMI, the ground state progressively shifts from the FM state, to a multi-$q$  state. We set the boundary of the FM phase at the limit where the FFT of the relaxed magnetic texture exhibits peaks away from $\overline{\Gamma}$.  The phase transition occurs at a scaling factor between 3 and 4 in FGT$_2$ and FGTS, and between only 2 and 3 in FGTSe, i.e., experimentally realistic values.
Example spin configurations in the second phase are shown in Figs.~\ref{fig:3q_spins}(a, c, e) for each system. In this phase, the DMI energy remains larger than in the $1q$ state, but the out-of-plane  components  of the magnetization lower the exchange energy (and potentially the in-plane DMI energy) compared to the $1q$ state and stabilize the $3q$ configuration. In the three systems, Fe$_3$ spins point uniformly out-of-plane. Interestingly, the states relaxed in FGT$_2$  yield 3 peaks of different amplitudes in the FFT: a main peak of 100\% intensity in $(q_x,q_y)=(\pm0.51,\pm0.03)$, a second peak at 65\% intensity in $(\pm0.50,\pm0,59)$, and a third peak at 25\% intensity in $(\pm0.50,\mp0.59)$. We refer to this state as asymmetric $3q$ state, with breaking of the C$_3$ rotation symmetry.
In the Janus MLs, we consistently find $3q$ states with $q=q_1=q_2=q_3 \approx 0.667$ and restored C$_3$ symmetry. 

We define a third phase when the $3q$ state minimizes the DMI energy, and no longer the exchange. The transition occurs at scaling factor between 6 and 7 in FGT$_2$, 7 and 8 in FGTSe, and 5 and 6 in FGTS.
The corresponding spin configurations resemble the DMI ground states shown in Figs.~\ref{fig:dmi_states_shellresolved} and ~\ref{fig:janus_dmi_states_shellresolved}. They correspond to the AFM ``nanoskX" in FGT$_2$ (Fig.~\ref{fig:3q_spins}(b)), the planar $3q$ state in FGTSe  (Fig.~\ref{fig:3q_spins}(d)), and a FM ``nanoskX" in FGTS (Fig.~\ref{fig:3q_spins}(f)). The latter is similar
to the state generated by Eq.~(\ref{eq:skx}) with $q=-2/3$, but with weak out-of-plane components of the Fe$_{1,2}$ spins.

\section{Discussion and conclusion}
In this work, we studied the effect of the out-of-plane DMI components in 2D FGTX (X=Te, Se, S) monolayers, namely a pristine FGT$_2$ monolayer, and two Janus monolayers FGTSe and FGTS, in which nonvanishing in-plane DMI is realized by breaking of the out-of-plane inversion symmetry. We used the extended Heisenberg Hamiltonian parametrized by DFT calculations in atomistic simulations, and also considered the same parameters scaled down according to DMFT, which yield a Curie temperature closer to experimental reports in FGT$_2$ monolayers. 

The energy dispersion of both out-of-plane and in-plane rotating spin spirals revealed that the DMI favors noncollinear states at the edge of the first BZ, characterized by a large exchange energy, and is therefore too weak to stabilize noncollinear ground states. 

By exploring the shell-resolved ground state of the sole DMI in the three systems, we found that the frustration of the OOP components tends to favor atomic-scale $3q$ magnetic textures over the typical $1q$ states, consisting of three wave vectors oriented 120 degrees from each other with an amplitude coinciding with the $\overline{K}$ points of the 1st BZ.  When first-neighbor in-plane DMI components are also considered, the relaxed textures are reminiscent of AFM nanoskyrmion lattices. 

Finally, owing to the possibility of tuning the DMI in 2D magnets via applied strain or electric field, we considered the effect of scaling the DMI components on the magnetic ground states. In the three systems, we found that $3q$ states are favored when DMI is enhanced by a factor between 2 and 4, which falls within the range of experimentally realistic values. We identified a first phase at low DMI that is exchange dominated, where nonplanar $3q$ states are realized. A second, DMI-dominated phase occurs at larger DMI, where a planar $3q$ state is relaxed in FGTSe, while we recovered the  peculiar (AFM) nanoskyrmion  lattices in FGT$_2$ and FGTSe.

Although we refer to such textures as ``nanoskyrmions" lattices, the relaxed entities are too small to carry a unit of topological charge. Unlike in a skyrmion lattice, they are also not separated by energy barriers, as the full texture tends to spontaneously relax from any perturbation of the collinear state, reminiscent of similar nanoscale multi-$q$ states stabilized by higher-order interactions~\cite{desplat2023eigenmodes}. 

 While these states lack a proper topological charge, their atomic size implies that they are rapidly traversed by conduction electrons, which may experience a spin flip, rather than the adiabatic spin rotation associated with a Berry phase. In this case, the non-adiabatic Hall effect may be observed, which is independent of topological charge~\cite{denisov2016electron,denisov2017nontrivial}. Our work thus adds another layer to the list of exotic spin physics already reported in FGT.

\begin{acknowledgments}
We graciously thank Libor Vojáček, Aur\'elien Manchon, Dongzhe Li, and Matthieu Jamet for fruitful discussions.
This work was supported by the FLAG-ERA grant MNEMOSYN, by the Scientific and Technological Research Council of Türkiye (TUBITAK) under project no. 221N400. R.C. acknowledges support from TUBITAK through the 2214-A International Research Fellowship Programme for PhD Students. Y. Mogulkoc acknowledges The Turkish Academy of Sciences – Outstanding Young Scientists Award Program (TUBA-GEBIP) for partial funding of the research. Additional support was provided by a
France 2030 government grant managed by the French
National Research Agency PEPR SPIN ANR-22-EXSP
0009 (SPINTHEORY). Computational resources
were provided by GENCI--IDRIS (Grant No.\ 2026-A0190912036) and by TUBITAK ULAKBIM High Performance and Grid Computing Center (TRUBA
resources). 
\end{acknowledgments}

\bibliography{references}

\end{document}